\documentclass[12pt,letterpaper]{JHEP3}
\usepackage{amsmath,amssymb,amsfonts,xspace}
\usepackage{epsfig}

\numberwithin{equation}{section}

\def\varpi{{\bf z}}

\def\sign{{\rm sign}}

\def\Im{\,{\rm Im}\,}
\def\Re{\,{\rm Re}\,}
\def\Ei{\,{\rm Ei}}

\def\({\left(}
\def\){\right)}
\def\[{\left[}
\def\]{\right]}
\def\hf{{1\over 2}}

\renewcommand{\d}{\mathrm{d}}
\newcommand{\de}{\mathrm{d}}

\newcommand{\I}{\mathrm{i}}
\newcommand{\e}{\mathrm{e}}
\newcommand{\cL}{\mathcal{L}}

\def\vrh{\varrho}

\newcommand{\p}{\partial}

\newcommand{\half}{\frac{1}{2}}

\newcommand{\cF}{\mathcal{F}}
\newcommand{\cV}{\mathcal{V}}

\newcommand{\cS}{\mathcal{S}}

\newcommand{\cK}{\mathcal{K}}
\newcommand{\cM}{\mathcal{M}}

\newcommand{\cN}{\mathcal{N}}

\newcommand{\CX}{\mathcal{X}}
\newcommand{\cR}{\mathcal{R}}

\DeclareSymbolFont{AMSa}{U}{msa}{m}{n}
\DeclareSymbolFont{AMSb}{U}{msb}{m}{n}
\DeclareMathSymbol{\fieldR}{\mathalpha}{AMSb}{"52}

\newcommand{\N}{{\mathcal N}}
\newcommand{\kahler}{{K\"ahler}\xspace}
\newcommand{\hk}{{hyperk\"ahler}\xspace}
\newcommand{\qk}{{quaternionic-K\"ahler}\xspace}

\newcommand{\cZ}{\mathcal{Z}}
\newcommand{\cI}{\mathcal{I}}
\newcommand{\cO}{\mathcal{O}}
\newcommand{\cH}{\mathcal{H}}
\newcommand{\cU}{\mathcal{U}}

\newcommand{\pa}{\partial}
\newcommand{\nn}{\nonumber}

\newcommand{\IR}{\mathbb{R}}
\newcommand{\IC}{\mathbb{C}}
\newcommand{\IZ}{\mathbb{Z}}

\newcommand{\Tr}{\mbox{Tr}}

\newcommand{\tzeta}{\tilde\zeta}

\newcommand{\txi}{\tilde\xi}

\newcommand{\CP}{\IC P^1}

\def\bea{\begin{eqnarray}}
\def\eea{\end{eqnarray}}
\def\be{\begin{equation}}
\def\ee{\end{equation}}
\def\ba{\begin{align}}
\def\ea{\end{align}}
\def\bse{\begin{subequations}}
\def\ese{\end{subequations}}

\def\bi{\bar \imath}
\def\bj{\bar \jmath}
\def\ba{\bar a}

\def\bz{\bar z}
\def\bY{\bar Y}

\def\bG{ \bar G }
\def\bW{ \bar W}

\def\bF{\bar F}

\newcommand{\CA}{{\cal{A}}}
\newcommand{\CB}{{\cal{B}}}
\newcommand{\CC}{{\cal{C}}}
\newcommand{\CD}{{\cal{D}}}

\newcommand{\CI}{{\cal{I}}}
\def\rf{r^{\flat}}
\def\ze{\zeta}

\def\ui#1{^{[#1]}}
\def\di#1{_{[#1]}}

\def\txii#1{{\tilde\xi}^{[#1]}}
\def\ai#1{{\alpha}^{[#1]}}
\def\nui#1{\nu_{[#1]}}
\def\xii#1{\xi_{[#1]}}
\def\Sij#1{S^{[#1]}}

\def\Hij#1{H^{[#1]}}

\def\Hpij#1{\Hij{#1}_{\scriptscriptstyle{\smash{(1)}}}}

\def\hHpij#1{\hHij{#1}_{\scriptscriptstyle{\smash{(1)}}}}

\newcommand{\Li}{{\rm Li}}

\def\Kkl{\cK_{\gamma}}
\def\Wkl{W_{\gamma}}
\def\bWkl{\bar W_{\gamma}}

\def\bWkl{\bW_{\gamma}}
\def\Thkl{\Theta_{\gamma}}
\def\Ikl{\cI_{\gamma}}
\def\Xikl{\Xi_{\gamma}}
\def\epskl{\epsilon_{\gamma}}

\def\Xint#1{\mathchoice
{\XXint\displaystyle\textstyle{#1}}%
{\XXint\textstyle\scriptstyle{#1}}%
{\XXint\scriptstyle\scriptscriptstyle{#1}}%
{\XXint\scriptscriptstyle\scriptscriptstyle{#1}}%
\!\int}
\def\XXint#1#2#3{{\setbox0=\hbox{$#1{#2#3}{\int}$}
\vcenter{\hbox{$#2#3$}}\kern-.5\wd0}}

\def\dashint{\Xint-}

\newcommand{\cwarrow}{\text{\Large$\curvearrowright$}}
\newcommand{\ccwarrow}{\text{\Large$\curvearrowleft$}}

\def\muh{\mu}
\def\hSij#1{S^{[#1]}}
\def\hHij#1{H^{[#1]}}
\newcommand{\hU}{\hat{\mathcal{U}}}
\newcommand{\hCX}{\mathcal{X}}
\def\hf{f}
\def\hH{H}
\def\cij#1{c}
\def\ci#1{c}
\def\hn{n}
\def\hnkl{n_{\gamma}}

\title{D-instantons and twistors}

\preprint{LPTA/08-073, ITP-UU-08-75,\\SPIN-08-58, IPhT-T08/203}

\author{Sergei Alexandrov$^1$, Boris Pioline$^{2}$,
Frank Saueressig$^3$, Stefan Vandoren$^4$
\\
$^1$ {\it Laboratoire de Physique Th\'eorique \&
Astroparticules, CNRS UMR 5207, \\
Universit\'e Montpellier II, 34095 Montpellier Cedex 05, France}\\

$^2$ {\it Laboratoire de Physique Th\'eorique et Hautes
Energies, CNRS UMR 7589, \\
Universit\'e Pierre et Marie Curie,
4 place Jussieu, 75252 Paris cedex 05, France} \\

$^3$ {\it Institut de Physique Th\'eorique,
 CEA, IPhT, F-91191 Gif-sur-Yvette, France\\
CNRS, URA 2306, F-91191 Gif-sur-Yvette, France}\\

$^4$ {\it   Institute for Theoretical Physics and
           Spinoza Institute,
           Utrecht University,
           Leuvenlaan 4,
           3508 TD Utrecht,
           The Netherlands
           }

\vspace*{2mm} {\tt e-mail: \email{alexandrov@lpta.univ-montp2.fr},
\email{pioline@lpthe.jussieu.fr}, \email{frank.saueressig@cea.fr},
\email{S.J.G.Vandoren@uu.nl}} \vspace*{-3mm}

}

\abstract{
Finding the exact, quantum corrected metric on the hypermultiplet
 moduli space in Type II string compactifications on Calabi-Yau threefolds
 is an outstanding open problem. We address this issue by relating
 the quaternionic-K\"ahler metric on the hypermultiplet moduli space to the
 complex contact geometry on its twistor space. In this framework,
 Euclidean D-brane instantons are captured by
 contact transformations between different patches.
 We derive those by recasting the previously known A-type D2-instanton
 corrections in the language of contact geometry, covariantizing the result
 under electro-magnetic duality, and using mirror symmetry.
 As a result, we are able to express the effects of all D-instantons in Type II
 compactifications  concisely as a sum of  dilogarithm functions. We conclude with
 some comments on the relation to microscopic degeneracies
 of four-dimensional BPS black holes and to the wall-crossing formula
 of Kontsevich and Soibelman, and on the form of the yet unknown NS5-brane instanton
 contributions.
}

\begin{document}

%%%%%%%%%%%%%%%%%%%%%%%%%%%%%%%%%%%%%%%%%%%%%%%%%%%%%%%%%%%%%%%%%%%%%%%%%%
\section{Introduction}
%%%%%%%%%%%%%%%%%%%%%%%%%%%%%%%%%%%%%%%%%%%%%%%%%%%%%%%%%%%%%%%%%%%%%%%%%%

Understanding non-perturbative corrections to the moduli space of hypermultiplets
in $\cN=2$ supersymmetric string vacua in $D=4$ and $D=3$ dimensions is an outstanding open
problem with a host of possible applications. Firstly, it would provide new checks of
heterotic/Type II string duality, which has mainly been tested in the vector multiplet
sector \cite{Aspinwall:1998bw}. Secondly, it may yield new insights on geometric
invariants of Calabi-Yau (CY) two- or threefolds, governing the contributions of Euclidean D-brane
and NS5-brane instantons in Type II strings, M2 and M5-brane
instantons in M-theory \cite{Becker:1995kb,Becker:1999pb}, or worldsheet instantons in heterotic
strings on $K_3$ \cite{Witten:1999fq,Halmagyi:2007wi}. Thirdly, it may provide a very useful packaging
of the BPS black hole degeneracies in four dimensions, via their relation
to $D=3$ BPS instantons \cite{Gunaydin:2005mx}.
Finally, it may be of practical use for phenomenological model building,
since the scalar potential in gauged supergravity typically depends on the metric on the
hypermultiplet branch, see e.g.
\cite{Kachru:2004jr,Davidse:2005ef,Saueressig:2005es,Grimm:2007xm,Looyestijn:2008pg}.

Contrary to the vector multiplet sector, where the relevant special \kahler (SK)
metrics can be obtained from a holomorphic prepotential, a major
difficulty in attacking this problem has been the lack of a convenient
parametrization of the \qk (QK) metrics on the hypermultiplet moduli space.
Recently, it has become clear that twistor techniques \cite{Penrose:1972ia,MR506229,
MR664330,MR1096180} are a powerful and practical tool for addressing this problem.
The relation of these mathematical constructions to the projective superspace
techniques developed in the physics literature in the context of
$\cN=2$ supersymmetric sigma models was gradually
understood in a series of works \cite{Hitchin:1986ea,Ivanov:1995cy,deWit:2001dj,
Alexandrov:2008ds,Lindstrom:2008gs,Alexandrov:2008nk}.

In particular, via Swann's construction \cite{MR1096180}
and the superconformal quotient construction \cite{deWit:2001dj},
QK (non-\kahler) manifolds $\cM$ in $4d$ real dimensions
are locally in one-to-one correspondence with
$4d+4$ dimensional \hk cones (HKC) $\cS$, i.e. HK manifolds with an isometric $SU(2)$
action and a homothetic Killing vector.
The HK metric on $\cS$ can be obtained from the
complex symplectic structure on its
twistor space $\cZ_\cS$, which in turn may be encoded in complex symplectomorphisms
relating different locally flat patches \cite{Alexandrov:2008ds}.
When $\cS$ is a HKC, the complex symplectic structure on $\cZ_\cS$ is homogeneous
and descends to a complex contact structure
on the twistor space $\cZ$ of $\cM$ \cite{MR664330}. The latter
may be described by complex contact transformations across different locally flat
patches \cite{Alexandrov:2008nk}.
Thus, the metric of a QK manifold can be encoded in a family of holomorphic functions
on $\cZ$ subject to consistency relations,  reality conditions and gauge equivalence.
This allows to by-pass the HKC $\cS$ and
its twistor space $\cZ_\cS$ altogether, even though the connection
to projective superspace is most obvious from the viewpoint of $\cZ_\cS$.

In the absence of isometries, computing the actual
QK metric on $\cM$ (or HKC metric on $\cS$) is in general
difficult, as it requires determining the real (contact) twistor lines.
It is however amenable to systematic approximation schemes when $\cM$
is a small deformation of a well understood QK manifold.
In particular, in \cite{Alexandrov:2008ds,Alexandrov:2008nk} we have given a general
formalism for  linear
perturbations of toric\footnote{A $4d$-dimensional HK
manifold is toric if it has $d$ commuting tri-holomorphic isometries.
A $4d$-dimensional QK
manifold is toric if it has $d+1$ commuting isometries. Both
 of these cases are covered by the Legendre
transform construction \cite{Karlhede:1984vr,Hitchin:1986ea}.}
HK and QK manifolds.

The projective superspace description of the hypermultiplet
moduli space at tree level in Type II compactifications
has been worked out in \cite{Rocek:2005ij,Rocek:2006xb}
(see also \cite{Berkovits:1995cb,Berkovits:1998jh} for some prescient work), and the one-loop
correction was incorporated in \cite{Robles-Llana:2006ez,Alexandrov:2007ec}, generalizing
earlier results \cite{Antoniadis:1997eg,Gunther:1998sc,Antoniadis:2003sw}.
Arguments for the absence of perturbative corrections beyond one-loop were given in
\cite{Gunther:1998sc,Antoniadis:2003sw,Robles-Llana:2006ez}.
More recently, by studying the fate of the
worldsheet instanton corrections
under S-duality,
the authors of \cite{RoblesLlana:2006is} were able to
compute the D(-1) and D1-instanton corrections to the hypermultiplet metric
in Type IIB string theory compactified on a CY threefold  $Y$.
Under mirror symmetry, these corrections translate into Euclidean D2-brane
instantons (or M2-branes in M-theory \cite{Becker:1995kb})
wrapped on special Lagrangian submanifolds
of the mirror CY threefold $X$ \cite{Saueressig:2007dr,RoblesLlana:2007ae},
recovering in particular the analysis of \cite{Ooguri:1996me} in the conifold limit.
However, these results do not include all D2-instanton contributions, since D(-1) and
D1-instantons are mirror symmetric to D2-branes wrapping A-type cycles only, where
 A- and B-cycles refer to a symplectic basis of
 $H_3(X)$ adapted to the point
of maximal unipotent monodromy (sometimes referred to as
the large complex structure limit); neither do they include NS5-brane instantons
(or M5-branes in M-theory).

The reason for restricting to
A-type D2-instantons (or D(-1) and D1-instan\-tons on the Type IIB side) is
that standard projective superspace
techniques rely on the existence of $d+1$ commuting
continuous isometries, which allow to dualize
all hypermultiplets into tensor multiplets. Generic instanton contributions preserve only
a  discrete subgroup of the continuous
isometries, and the resulting metric falls outside the class of metrics obtainable by the
Legendre transform method \cite{Karlhede:1984vr,Hitchin:1986ea}.\footnote{HK and QK metrics
obtainable from the generalized Legendre transform \cite{Lindstrom:1987ks}
have generically no isometries, but
still possess a higher rank Killing tensor; the metric on the universal hypermultiplet
in the presence of NS5-brane instantons has been argued to fall in this class
\cite{Anguelova:2004sj}.} However, the general
construction of HK (resp., QK) manifolds from complex symplectic (resp., contact)
manifolds with a compatible real structure remains valid. It may well be feasible
to determine the complex symplectic (or contact) structure on the twistor space
exactly, e.g. by specifying a set of complex symplectomorphisms
(or contact transformations) between
different locally flat patches, even
if the exact HK (QK) metric remains out of reach.

This strategy was applied recently to the case of $D=3, \cN=4$
supersymmetric gauge theories in 2+1 dimensions, obtained from compactifying
$D=4, \cN=2$ supersymmetric gauge theories on a circle \cite{Gaiotto:2008cd}.
In this case, the moduli space is HK, and receives instanton corrections from
4D BPS solitons winding around the Euclidean circle.
The elementary symplectomorphism induced by such a soliton
can be computed unambiguously in field theory, and
a natural way to combine contributions from mutually non-local
solitons suggests itself \cite{Gaiotto:2008cd}. While the (indexed)
one-particle BPS spectrum jumps across lines of marginal stability (LMS),
the multi-particle BPS spectrum and more generally any physical observable
should be  smooth across the LMS. Indeed, the authors of \cite{Gaiotto:2008cd}
show that the HK metric derived from the symplectic structure is
regular across the LMS, provided the change in the one-particle
BPS spectrum satisfies constraints
identical in form to the  wall-crossing formula for ``generalized Donaldson-Thomas
invariants'' \cite{ks} (see \cite{moorelec}
for a physics discussion of this formula).

In this paper, we initiate a similar study in the context of $\cN=2$ supergravity
in four dimensions, and obtain the
contributions of all A-type and B-type D-instantons (with vanishing NS5-brane charge)
to the hypermultiplet metric in Type II compactifications.
Following the roadmap laid out in \cite{RoblesLlana:2007ae},
we proceed by covariantizing the known A-type contributions
under electric-magnetic duality and using mirror symmetry (in the process,
we clarify the action of the latter on the Ramond-Ramond (RR) potentials, at
least in the large volume limit).
We work in the ``leading instanton approximation", treating the
D-instantons as linear perturbations around the one-loop corrected
geometry of \cite{Robles-Llana:2006ez}\footnote{It is possible in principle to treat the
A-type instantons exactly as in \cite{RoblesLlana:2006is,RoblesLlana:2007ae,Saueressig:2007gi}
and the
B-type instantons as linear perturbations. While this constitutes a valid approximation in the
limit of large complex structures or \kahler classes,  this approach is not directly useful
as it breaks electric-magnetic covariance.} using the formalism developed
in \cite{Alexandrov:2008ds,Alexandrov:2008nk}.
In particular,  we obtain the instanton corrected twistor lines \eqref{xiqlineB} and
contact potential \eqref{phiinstfull} (related to the \kahler potential
on $\cZ$  via \eqref{Knuflat}), and show that the D-instanton effects
can be concisely summarized in a holomorphic function
\eqref{prepH}, expressed as a sum of  dilogarithms, controlling the
deformation of the  complex contact structure on $\cZ$.

Our results above should  really be viewed as
a parametrization of the instanton corrected
hypermultiplet metric. While the coefficients
of mixed A/B-type contributions are in principle new geometric invariants
of the CY threefold, we do not know how to compute
them, although they should be obtainable from the  dual heterotic sigma
model \cite{Aspinwall:1998bw,Witten:1999fq,Halmagyi:2007wi}.
The analogy with the results of \cite{ks},
most notably the appearance of the dilogarithm function,
strongly suggests that these invariants should be identified with the
generalized Donaldson-Thomas invariants defined in \cite{ks}.
It is possible that using the wall-crossing formula of \cite{ks}, possibly combined
with some automorphy requirement, one may be able to fix these invariants
completely. By reduction from 4D to 3D and T-duality
on the circle, the same invariants should
determine the exact micro-state degeneracies of 4D black holes, as we further discuss in
Section \ref{bhsec}.

This paper is organized as follows. $\bullet$ In Section \ref{sec_projdescr},
we summarize the twistorial
description of general QK manifolds and the linear deformations of
toric QK metrics. $\bullet$  In Section \ref{sec_perturb} we describe the hypermultiplet
moduli space at the perturbative level, and discuss the action of S-duality
and mirror symmetry.
$\bullet$  In Section \ref{sec:Dinst}, we formulate the A-type instanton corrections
in terms of the contact geometry on the twistor space,
and use electric-magnetic duality and mirror symmetry to obtain the effect of
mixed A and B-type instantons in the leading instanton approximation.
We derive the instanton corrected twistor lines and \kahler potential,
and suggest a construction of the instanton corrected twistor space
beyond  the leading instanton approximation in the spirit of \cite{Gaiotto:2008cd}.
$\bullet$  In Section \ref{sec_disc},
we relate D-instanton corrections to the 4D hypermultiplet moduli space
to corrections to the 3D vector multiplet moduli space  induced by
4D BPS black holes, discuss the usefulness of this approach in
incorporating the moduli dependence of the black hole
micro-state degeneracies, comment on possible
relations to the generalized Donaldson-Thomas invariants of \cite{ks}
and  on the form of NS5-instanton contributions.
$\bullet$  In Appendix \ref{ap_GMN}, we revisit the construction of
the twistor space of the Ooguri-Vafa metric
discussed in \cite{Gaiotto:2008cd}, and extend it
to provide a rigorous construction of the twistor space
of the hypermultiplet branch in the leading instanton approximation.

%------------------------------------------------------------------------
\section{QK spaces, twistors and  contact geometry}
\label{sec_projdescr}
%------------------------------------------------------------------------

In this section, we give a streamlined summary of our recent work
\cite{Alexandrov:2008nk} on the twistor approach to QK geometry,
retaining only the information relevant for the twistor space $\cZ$ of $\cM$,
and with a few changes of notations in order to avoid
cluttering.\footnote{In particular, we drop the ``hat'' on all symbols
$\hat f_{ij}, \hat\CX\ui{i}, \hat S^{[ij]},
\hat H^{[ij]}$, replace the index $\flat$ by the subscript $\alpha$,
and rephrase all contact transformations in terms of the variable
$\ai{i}$ rather than $\txii{i}_\flat$. This removes
the $c_I$ dependence from the contact transformations
(2.71) and (5.23) in \cite{Alexandrov:2008nk}. Moreover, since
only the anomalous dimensions in the patch $\cU_\pm$ play
a role in the present construction, we denote
$c_\Lambda\equiv c^{[+]}_\Lambda=-c^{[-]}_\Lambda$, $c_\alpha\equiv c^{[+]}_\flat=-c^{[-]}_\flat$.}
Further mathematical details can be found, e.g., in \cite{MR664330,MR1327157}.

%------------------------------------------------------------------------
\subsection{General \qk manifolds}
\label{sect:2.1}
%------------------------------------------------------------------------

A QK manifold $\cM$ is a $4d$-dimensional Riemannian manifold whose
holonomy is contained in $USp(d)\times SU(2)$.
It admits a quaternionic structure, which locally yields
three almost complex structures satisfying the algebra of the unit quaternions.
$\cM$ is conveniently described by
its twistor space $\cZ$,  a $\CP$ bundle over $\cM$, whose connection is given
by the $SU(2)$ part $\vec p$ of the Levi-Civita connection on $\cM$.
$\cZ$ admits a canonical (integrable) complex structure and a \kahler-Einstein
metric \cite{MR664330}. The latter can be written as
\begin{equation}\label{Z-metric}
\de s^2_{\cZ}=
\frac{|D\varpi|^2}{(1+\varpi{\bar \varpi})^2}+\frac{\nu}{4}\,\de s_{\cal M}^2\, .
\end{equation}
Here $\varpi$ is a complex coordinate\footnote{Note, however, that the projection
$\cZ \to \CP$, $(u,\bar u)\mapsto \varpi$ is not holomorphic.}
on $\CP$, $D\varpi$ is the canonical $(1,0)$ form
\be
\label{defdz}
D\varpi \equiv  \de\varpi + p_+ -\I p_3 \,\varpi + p_-\, \varpi^2\, ,
\ee
with  $p_- = (p_+)^*$, $p_3 = (p_3)^*$
under complex conjugation, and
$\nu = R/(4d(d+2))$ is a numerical constant which sets the constant curvature $R$ of $\cM$.

The kernel of $D\varpi$ endows  $\cZ$ with a complex contact
structure \cite{MR664330,MR1327157}
(see e.g. \cite{MR2194671} for a general introduction to contact geometry).
The latter can be represented by
a set of holomorphic one-forms $\hCX\ui{i}$ defined on an open covering $\hU_i$ of $\cZ$,
such that the holomorphic top form $\hCX\ui{i} \wedge (\de\hCX\ui{i})^d$ is nowhere vanishing.
On each patch, $\hCX\ui{i}$ is proportional to $D\varpi$,
\be
\label{contact}
\hCX\ui{i} = 2\, e^{\Phi\di{i}} \frac{D\varpi}{\varpi}\, ,
\ee
where $\Phi\di{i}\equiv\Phi\di{i}(x^\mu,\varpi)$
is a function on $\hU_i\subset\cZ$ which we refer to as the ``contact potential".
It is holomorphic along the $\CP$ fiber, defined up to an additive holomorphic
function on $\hU_i$, and chosen such that the right-hand
side of \eqref{contact} is a holomorphic (i.e. $\bar\pa$-closed) one-form.
The reality constraint
\be
\overline{\tau(\hCX\ui{i})}= - \hCX\ui{\bi}\ , \label{relcontact}
\ee
where $\tau$ is the antipodal map acting as $\tau: \varpi\to-1/\bar\varpi$
on $\CP$ and relating the two patches
$\hU_i$ and $\hU_{\bi}$, requires that
\be
\overline{\tau(\Phi\di{i})}=\Phi\di{\bi}\, .
\ee

The real part of $\Phi\di{i}$ provides a \kahler potential
for the \kahler-Einstein metric on $\cZ$ in the patch $\hU_i$,
\be
\label{Knuflat}
K_{\cZ}\ui{i} = \log\frac{1+\varpi\bar \varpi}{|\varpi|}
+ \Re\Phi\di{i}(x^\mu,\varpi)\, .
\ee
One way to compute the metric on $\cZ$ and $\cM$ would be to express
\eqref{Knuflat} in terms of complex  coordinates on $\cZ$, which is in general
difficult. Fortunately, we shall be able to obtain the metric without knowing
this change of coordinates. Note that the metric \eqref{Z-metric} now rewrites as
\be\label{twistor-metric}
\de s^2_{\cZ}=\frac14 \left( e^{-2K_\cZ} |\hCX|^2\,
+\nu \, \de s_{\cal M}^2\right) \, ,
\ee
consistently with \cite{deWit:2001dj} where \eqref{twistor-metric} was written
in a particular gauge.

By a simple extension of Darboux's theorem \cite{MR1327157}, one may choose
the open covering $\hU_i$ such that on each patch $\hCX\ui{i}$ takes the canonical form
\be
\label{contf}
\hCX\ui{i}= \de\ai{i} + \xii{i}^\Lambda \, \de \txii{i}_\Lambda\, .
\ee
Here $\xii{i}^\Lambda, \txii{i}_\Lambda, \ai{i}$ ($\Lambda=0,\dots,d-1$)
are local complex coordinates on $\cZ$, smooth throughout the patch $\hU_i$.
Moreover,  these coordinates may be chosen to satisfy the reality conditions
\be
\label{rexixit}
\overline{\tau(\xii{i}^\Lambda)} = \xii{\bar \imath}^\Lambda\, ,
\qquad
\overline{\tau(\txii{i}_\Lambda)} = -\txii{\bi}_\Lambda  
\, ,
\qquad
\overline{\tau(\ai{i})} = -\ai{\bi} 
\, .
\ee

While the form \eqref{contf} can always be achieved locally by a choice of complex coordinates,
the global complex contact structure on $\cZ$ is encoded in the set of
{\it complex contact} transformations which relate  the two systems of complex coordinates
$(\xii{i}^\Lambda, \txii{i}_\Lambda, \ai{i})$ and $(\xii{j}^\Lambda, \txii{j}_\Lambda, \ai{j})$ on the
overlap $\hU_i\cap \hU_j$. Complex contact transformations are holomorphic
transformations which obey
\be
{\hCX\ui{i}}={\hf}_{ij}^2 \, {\hCX\ui{j}} \, ,
\ee
for some nowhere vanishing holomorphic function ${\hf}_{ij}^2$ on $\hU_i\cap \hU_j$.
They can generally be represented\footnote{One way to see this is
to ``symplectize" the contact form, i.e. introduce an extra local complex variable $\nui{i}^\alpha$
and consider the homogeneous symplectic form
$\Omega\ui{i}=\de(\nui{i}^\alpha \hCX\ui{i})$.}
 by a holomorphic function
$\hSij{ij}(\xii{i}^\Lambda,\txii{j}_\Lambda,\ai{j})$ of the ``initial position''
$\xii{i}^\Lambda$, ``final momentum" $\txii{j}_\Lambda$ and ``final action" $\ai{j}$
such that
\be\label{xitrafo}
\begin{split}
\xi\di{j}^\Lambda &=  \hf_{ij}^{-2} \, \pa_{\txii{j}_\Lambda} \hSij{ij}\, ,
\qquad\quad\quad\
\txii{i}_\Lambda = \pa_{\xii{i}^\Lambda} \hSij{ij}\, ,
\\
\ai{i} &= \hSij{ij} - \xii{i}^\Lambda \pa_{\xii{i}^\Lambda} \hSij{ij}\ ,
\qquad
\hf_{ij}^{2} = \pa_{\ai{j}} \hSij{ij}\, ,
 \end{split}
\ee
on $\hU_i\cap \hU_j$. In particular, the contact potentials satisfy
\be
\label{Phiij}
e^{\Phi\di{i}} =  \hf_{ij}^{2} \,e^{\Phi\di{j}} \, .
\ee
As explained in  \cite{Alexandrov:2008ds,Alexandrov:2008nk},
the functions $\hSij{ij}$ are subject to several conditions: (i) consistency conditions ensuring
that the contact transformations compose properly on the triple overlap
$\hU_i\cap \hU_j\cap \hU_k$,
\be
\hSij{ij}(\xii{i}^\Lambda,\txii{j}_\Lambda,\ai{j})=
\left\langle \hSij{ik}(\xii{i}^\Lambda,\txii{k}_\Lambda,\ai{k})
-
\lambda \left( \ai{k}+ \xii{k}^\Lambda \txii{k}_\Lambda
-\hSij{kj}(\xii{k}^\Lambda,\txii{j}_\Lambda,\ai{j})\right)
\right\rangle,
\label{compS}
\ee
where the bracket denotes extremization over $\xii{k}^\Lambda,\txii{k}_\Lambda,\ai{k}$ and the
Lagrange multiplier $\lambda$;
(ii) gauge equivalence  generated by holomorphic functions $T^{[i]}$ in each patch $\hU_i$,
\bea
\hSij{ij}(\xii{i}^\Lambda,\txii{j}_\Lambda,\ai{j})&\mapsto&
\left\langle
T^{[i]}(\xi_{[i]}^{\Lambda}, \txi^{\,'[i]}_\Lambda,\alpha^{'[i]})-
\lambda_1 \left ( \alpha^{'[i]} + \xi_{[i]}^{'\,\Lambda} \txi^{\,'[i]}_\Lambda
-\hSij{ij}( \xi_{[i]}^{'\Lambda}, \txi^{\,'[j]}_\Lambda,\alpha^{'[j]})
 \right)
\right.
\nonumber \\
&&\left.
+\lambda_2
\left( \alpha^{[j]} + \xi_{[j]}^{\Lambda} \txi^{[j]}_\Lambda
-T^{[j]}(\xi_{[j]}^{\Lambda}, \txi^{\,'[j]}_\Lambda,\alpha^{'[j]})
\right)\right\rangle\, ,
\label{gaugeT}
\eea
where the bracket denotes extremization over $\xi_{[i]}^{'\Lambda},\txi^{'[i]}_\Lambda,\alpha^{'[i]},
\xi_{[j]}^{'\Lambda}, \txi^{\,'[j]}_\Lambda,\alpha^{'[j]}$ and the Lagrange multipliers $\lambda_1,
\lambda_2$;
and (iii) reality conditions ensuring \eqref{relcontact},
\be
\overline{\tau(\hSij{ij})} = -\hSij{\bi\bj}\, .
\label{realcon}
\ee

A particularly important case occurs when $\hf_{ij}^{2}$ can
be chosen all equal to one. In this case, $\hSij{ij}$ reduces to
\begin{equation}
\hSij{ij}(\xi^\Lambda_{[i]},{\tilde \xi}_\Lambda^{[j]},\alpha^{[j]})=
\alpha^{[j]}+
\Sij{ij}(\xi^\Lambda_{[i]},{\tilde \xi}_\Lambda^{[j]})\, ,
\label{Sspe}
\end{equation}
where $\Sij{ij}$ is the generating function of a symplectomorphism of
the $(\xi^\Lambda,\txi_\Lambda)$ phase space, and $\hCX$ becomes
a global contact one-form with Reeb vector $\pa_{\ai{i}}$.
The contact potentials $\Phi\di{i}$ are  then all equal
to a single real function $\Phi(x^\mu)$, constant along
the $\CP$ fiber (as follows from \eqref{Phiij} and the requirement
of holomorphy in all patches). Toric QK manifolds discussed in Section \ref{sectoric} below
fall in this class, and so do hypermultiplet moduli spaces in the absence of NS5-brane
instantons, as discussed in Section \ref{linins}.

In order to construct the QK metric on $\cM$, one should first determine the ``contact twistor lines",
i.e. express the local complex coordinates  $\xii{i}^\Lambda, \txii{i}_\Lambda, \ai{i}$ on $\cZ$
in terms of the coordinates $x^\mu$ on the base and the coordinate $\varpi$ on the fiber.
In the patch $\hU_+$ around $\varpi=0$, the  coordinates must be
smooth up to specific singular terms compatible
with the form of \eqref{contact}  \cite{Alexandrov:2008nk},
\be\label{behap}
\begin{split}
\xii{+}^\Lambda &= \xii{+}^{\Lambda,-1}\, \varpi^{-1} + \xii{+}^{\Lambda,0}
+ \xii{+}^{\Lambda,1}\,  \varpi +\cO( \varpi^2)\, ,  \qquad
\\
\txii{+}_\Lambda &=  \ci{+}_\Lambda \log\varpi + \txi^{[+]}_{\Lambda,0}
+ \txi^{[+]}_{\Lambda,1}\, \varpi+\cO( \varpi^2)\, ,
\\
\ai{+} &= \ci{+}_\alpha  \log\varpi + \ci{+}_\Lambda \, \xii{+}^{\Lambda,-1} \,\varpi^{-1}
+ \alpha^{[+]}_0+ \alpha^{[+]}_1 \,\varpi+ \cO( \varpi^2)\, ,
\\
\Phi\di{+}&= \phi\di{+}^0 +  \phi\di{+}^1 \, \varpi +\cO( \varpi^2) \, .
\end{split}
\ee
Here the coefficients $\ci{+}_I$, with the index $I$ running over $\alpha,0,\ldots,d-1$,
are complex numbers called ``anomalous dimensions''.
As a result of the logarithmic singularity
in \eqref{behap}, the last two reality conditions in \eqref{rexixit}
pick up additive constants, which however do not affect the reality condition on $\hCX\ui{i}$.
Generically, all Laurent coefficients in \eqref{behap}
are determined from the lowest coefficients
$\xii{+}^{\Lambda,-1},\txi^{[+]}_{\Lambda,0},\alpha^{[+]}_0$
by imposing the gluing conditions
\eqref{xitrafo}, and parametrize the manifold $\cM$, up to
overall phase rotations of $\xii{+}^{\Lambda,-1}$.

The \qk metric $g$ on $\cM$ may then be recovered as follows (see \cite{Alexandrov:2008nk} for more
details and explicit examples). Firstly,
by expanding $e^{-\Phi_{[+]}}\hCX^{[+]}$ in \eqref{contf} around $\varpi=0$, and comparing
with \eqref{contact}, one may extract the $SU(2)$ connection
\be
\begin{split}\label{connection}
p_+ &=\frac12\, e^{-\phi\di{+}^0}
\left( \xi^{\Lambda,-1}_{[+]}  \de\txi^{[+]}_{\Lambda,0}
+ \ci{+}_\Lambda \de\xi^{\Lambda,-1}_{[+]}
\right)\, ,
\\
p_3 &= \frac{\I}{2}\, e^{-\phi\di{+}^0} \left( \de\alpha^{[+]}_0 +
\xi^{\Lambda,0}_{[+]}  \de\txi^{[+]}_{\Lambda,0} +
\xi^{\Lambda,-1}_{[+]}  \de\txi^{[+]}_{\Lambda,1}  \right)
- \I \phi\di{+}^1 p_+\, ,
\end{split}
\ee
and express the Laurent coefficients of the contact potential in terms of the
Laurent coefficients of the contact twistor lines,
\be
\label{contpot}
\begin{split}
e^{\phi\di{+}^0} =&  \frac12 \left( \xi^{\Lambda,-1}_{[+]} \txi^{[+]}_{\Lambda,1}
+ \ci{+}_\Lambda \xii{+}^{\Lambda,0} +   \ci{+}_\alpha \right)\, ,
\\
\phi\di{+}^1 = & \frac12 \,e^{-\phi\di{+}^0}
\left( \alpha^{[+]}_1 + 2 \xi^{\Lambda,-1}_{[+]}  \txi^{[+]}_{\Lambda,2}
+  \xi^{\Lambda,0}_{[+]}  \txi^{[+]}_{\Lambda,1} + \ci{+}_\Lambda \xi^{\Lambda,1}_{[+]}  \right)\, .
\end{split}
\ee
Subsequently, one expands the holomorphic one-forms
$\de\xii{+}^\Lambda,\ \de\txii{+}_\Lambda$ and $\de\alpha$
around $\varpi=0$ and projects them along the base $\cM$, producing
local one-forms on $\cM$ of Dolbeault type $(1,0)$  with respect to the
quaternionic structure $J_3$. A basis of these forms is given by
\be
\label{defPi}
\Pi^a =\xi_{[+]}^{0,-1} \cV^a - \xi_{[+]}^{a,-1} \cV^0\, ,
\qquad
\tilde\Pi_I = \xi_{[+]}^{0,-1}  \tilde{\cV}_I + \ci{+}_I \cV^0\, ,
\ee
where $a$ runs over $1,\ldots,d-1$, and
\be
\label{xi10}
\begin{split}
\cV^\Lambda&\equiv  (\de - \I p_3)  \xi_{[+]}^{\Lambda,-1}\, ,
\qquad
\tilde{\cV}_\Lambda \equiv  \de\txi^{[+]}_{\Lambda,0}
- \txi^{[+]}_{\Lambda,1} \, p_+  + \I \ci{+}_\Lambda  p_3 \, ,
\\
\qquad
\tilde{\cV}_\alpha& \equiv  \de\alpha^{[+]}_{0} - c_\Lambda  \de\xi_{[+]}^{\Lambda,0}
- (\alpha^{[+]}_{1} - \ci{+}_\Lambda  \xi_{[+]}^{\Lambda,1}) p_+  + \I\, \ci{+}_\alpha   p_3 \, ,
\end{split}
\ee

Then, one may compute the triplet of quaternionic 2-forms $\vec\omega$ from the
curvature of the $SU(2)$ connection
\be\label{su2curv}
{\rm d} \vec p + \frac{1}{2}\,\vec p \times
\vec p = \frac{\nu}{2}\,\vec \omega \, ,
\ee
where $\nu$ is a constant
related to the scalar curvature of $\cM$. In particular,
we have (without loss of generality we will set $\nu =2$ in the following)
\be
\omega_3 = {\rm d} p_3 + 2\I  p_+ \wedge p_- \, ,
\ee
and obtain the QK metric via $g = \omega_3 \cdot J_3$.

%------------------------------------------------------------------------
\subsection{Toric quaternionic-\kahler geometries}
\label{sectoric}
%------------------------------------------------------------------------

A particularly simple case occurs for toric QK manifolds, i.e. when the $4d$-dimensional
QK manifold $\cM$ admits $d+1$
commuting isometries. In this case, the moment maps associated to these
isometries \cite{MR872143} provide $d+1$
independent global $\cO(2)$ sections on the twistor space $\cZ$ of $\cM$, which can
be taken to be the complex coordinates $\xii{i}^\Lambda$ and the unit function.
Thus, on all  patches $\hU_i$,
$\xii{i}^\Lambda$ takes the form
\be
\label{gxi}
\xii{i}^\Lambda = \xi^\Lambda \equiv Y^\Lambda \varpi^{-1} + A^\Lambda - \bY^\Lambda \varpi\, .
\ee
Moreover, the $U(1)$ action corresponding to phase rotations of $\varpi$ can
be fixed by choosing $Y^0\equiv \cR$ to be real.
Supplemented by $d+1$ additional coordinates $B_I$ to be defined below,
$\cR, Y^a, \bY^a, A^\Lambda$ provide a convenient coordinate system on $\cM$.

On the overlap of two patches, the complex coordinates
$\xi^\Lambda, \txi_\Lambda, \alpha$
must now be related by a complex contact transformation which preserves $\xi^\Lambda$
and the unit function.
This restricts the generating function $\hSij{ij}$ to the form
\be
\label{Siso}
\hSij{ij} = \ai{j}
 + \xii{i}^\Lambda \,  \txii{j}_\Lambda
- \hHij{ij}( \xii{i}^\Lambda)\, ,
\ee
where $\hHij{ij}(\xi^\Lambda)$ is  a holomorphic function on ${\hU}_i\cap\hat{\cU}_j$.
The contact transformations \eqref{xitrafo} become
\bea
\label{contactmu}
\txii{i}_\Lambda &=&  \txii{j}_\Lambda - \pa_{\xi^\Lambda} \hHij{ij}\, ,
\qquad \ai{i} =  \ai{j} - \hHij{ij}
  +\xi^\Lambda\pa_{\xi^\Lambda} \hHij{ij} \, ,
\eea
and the transition function $\hf_{ij}^{2}$ is now equal to one.
The consistency conditions \eqref{compS}, gauge equivalence \eqref{gaugeT}
and reality conditions \eqref{realcon} translate into
\be
\label{consisth}
\hHij{ij}+ \hHij{jk} = \hHij{ik}\, ,
\qquad
\hHij{ij} \mapsto  \hHij{ij}+ T^{[i]}-T^{[j]}\, ,
\qquad
\overline{\tau(\hHij{ij})} = -\hHij{\bi\bj}\, .
\ee
We shall often abuse notation and define $\hHij{ij}$ away from the overlap $\hU_i \cap \hU_j$
(in particular when the two patches do not intersect) using analytic continuation and the
first equation in \eqref{consisth} to interpolate from $\hU_i$ to $\hU_j$. Ambiguities in the
choice of path can be dealt with on a case by case basis.

The gluing conditions \eqref{contactmu} are sufficient to determine $\txii{i}_\Lambda, \ai{i}$
uniquely, up to overall real constants $B_\Lambda, B_\alpha$ which provide the extra $d+1$ coordinates
mentioned above,
\bea
 \label{eqmuqh}
\txii{i}_\Lambda &=&
\frac{\I}{2}\, B_\Lambda+\frac12\sum_j\oint_{C_j}\frac{\de\varpi'}{2\pi\I\varpi'}\,
\frac{\varpi'+\varpi}{\varpi'-\varpi}\, \pa_{\xi^\Lambda}  \hHij{+j}(\xi (\varpi'))
+\cij{+}_\Lambda \log \varpi\, ,
\\
\ai{i} &=& \frac{\I}{2}\, B_\alpha+\frac12\sum_j\oint_{C_j}\frac{\de\varpi'}{2\pi\I \varpi'}\,
\frac{\varpi'+\varpi}{\varpi'-\varpi}\, \left[ \hH - \xi^\Lambda  \pa_{\xi^\Lambda} \hH
 \right]^{[+j]}
+\cij{+} _\alpha \log \varpi
+\cij{+}_\Lambda\(Y^\Lambda \varpi^{-1} + \bY^\Lambda \varpi\) \, .
\nn
\eea
Here $\varpi\in \cU_i$ and $C_j$ is a contour surrounding $\cU_j$, with $\cU_i$
denoting the projection of $\hat{\cU}_i$ to $\CP$. Eqs. \eqref{gxi} and
 \eqref{eqmuqh} exhibit the complex coordinates on $\cZ$ as
functions of the coordinates $(\cR, Y^a, \bY^a, A^\Lambda, B_I)$ on $\cM$
and of the complex coordinate $\varpi$ on $\CP$, and parameterize the ``contact twistor lines''.
Furthermore, in the toric case the potential $\Phi\di{i}(x^\mu,\varpi)\equiv\Phi(x^\mu)$
is independent of $\varpi$ and the same in all patches,
\be
\label{eqchi}
e^{\Phi}=\frac14 \, \sum_j\oint_{C_j}\frac{\de\varpi'}{2\pi\I \varpi'}
\(\varpi'^{-1} Y^{\Lambda}-\varpi' \bY^{\Lambda} \)\pa_{\xi^\Lambda}
{\hH^{[+j]}}(\xi(\varpi')) +\frac12\( \cij{+}_\Lambda  A^\Lambda +  \cij{+}_\alpha\) \, .
\ee
Note that due to consistency conditions \eqref{consisth}, the index $\scriptstyle [+]$ in
\eqref{eqmuqh}, \eqref{eqchi} can be replaced by any other patch index without affecting the result.

Let us mention also that for the purpose of expressing the metric on $\cM$,   it is sometimes
more convenient to trade the coordinate $\cR$ for the variable $\Phi$.
As we shall see below, this is natural for the hypermultiplet moduli space,
since the contact potential $\Phi$ is identified with the four-dimensional dilaton $\phi$.

%------------------------------------------------------------
\subsection{Linear deformations}
%------------------------------------------------------------
\label{subsec_lindef}

Deformations of a QK
manifold $\cM$ which preserve the QK property  are controlled by
the sheaf cohomology group
$H^1(\cZ,\cO(2))$ \cite{lebrun1988rtq,lebrun1994srp}. In practice, this means that they
correspond to infinitesimal perturbations of the complex contact structure obtained
by replacing
\be
\hHij{ij}(\xii{i}) \rightarrow \hHij{ij}(\xii{i}^\Lambda)
+ \hHpij{ij}(\xii{i}^\Lambda, \txii{j}_\Lambda,\ai{j})
\ee
in \eqref{Siso}, preserving the co-cycle conditions, reality conditions and modulo
local contact transformations as in \eqref{consisth} (where now all quantities are
functions of $(\xii{i}^\Lambda, \txii{j}_\Lambda,\ai{j})$)  \cite{Alexandrov:2008nk}. The
function $\hHpij{ij}$, holomorphic on $\hU_i\cap \hU_j$, corresponds to the contact Hamiltonian
(or moment map) of the infinitesimal contact transformation performed in gluing the
patches $\hU_i$ and $\hU_j$. As mentioned below \eqref{consisth}, we abuse notation
and consider $\hHpij{ij}$ even when the patches do not intersect.

In general (for $\txii{j}_\Lambda$
and $\ai{j}$-dependent $\hHpij{ij}$) the perturbations break the $d+1$ isometries, and
the position coordinate $\xii{i}^\Lambda$ is no longer a global $\cO(2)$ section. Indeed,
the contact transformations \eqref{xitrafo} become, to linear order in
the perturbation,
\be
\label{pert:symp}
\begin{split}
\xii{i}^\Lambda &=   \xii{j}^\Lambda - T_{[ij]}^\Lambda
\, ,
\qquad\quad
\txii{i}_\Lambda =  \txii{j}_\Lambda - \tilde{T}^{[ij]}_\Lambda
\, ,
\\
\ai{i} &=  \ai{j} - \tilde{T}^{[ij]}_\alpha\, ,
\qquad\
\Phi_{[i]}  =  \Phi_{[j]} - \p_{\ai{j}}\hHpij{ij}\, ,
\end{split}
\ee
where we denoted
\be\label{Tfct}
\begin{split}
T_{[ij]}^\Lambda \equiv &
- \p_{\txii{j}_\Lambda }\hHpij{ij}
+ \xii{i}^\Lambda \, \p_{\ai{j}}\hHpij{ij}
\, ,
%\\
\qquad
\tilde{T}^{[ij]}_\Lambda \equiv
 \p_{\xii{i}^\Lambda }
\left(  \hHij{ij} + \hHpij{ij} \right)
\, ,
\\
\tilde{T}^{[ij]}_\alpha \equiv &
 \left(  \hHij{ij}+ \hHpij{ij} \right)
- \xii{i}^\Lambda \p_{\xii{i}^\Lambda}
\left(  \hHij{ij} + \hHpij{ij} \right)
\, .
\end{split}
\ee
In Eq. \eqref{mucontact} below,
we interpret $(T_{[ij]}^\Lambda,\tilde{T}^{[ij]}_\Lambda,\tilde{T}^{[ij]}_\alpha)$
as the contact vector field derived from the contact Hamiltonian $\hHij{ij}+ \hHpij{ij}$.

In the linear approximation the arguments of $\hHpij{ij}$ can be taken to
be the unperturbed contact twistor lines defined in \eqref{gxi} and \eqref{eqmuqh}.
It is then straightforward to compute the correction to these unperturbed quantities,
\bea
\xii{i}^\Lambda(\varpi,x^\mu)& =& A^\Lambda +
\varpi^{-1} Y^\Lambda - \varpi \bY^\Lambda
+\frac12 \sum_j \oint_{C_j}\frac{\de\varpi'}{2\pi\I \varpi'}\,
\frac{\varpi'+\varpi}{\varpi'-\varpi}\, T_{[+j]}^\Lambda(\varpi')\, ,
\nonumber \\
\txi_\Lambda^{[i]}(\varpi,x^\mu)& = &\frac{\I}{2}\, B_\Lambda +
\half  \sum_j \oint_{C_j} \frac{\de \varpi'}{2 \pi \I \varpi'} \,
\frac{\varpi' + \varpi}{\varpi' - \varpi}
\, \tilde{T}_\Lambda^{[+j]}(\varpi')+  \ci{+}_\Lambda \log \varpi
\, ,
\label{txiqline}
\\
\ai{i}(\varpi,x^\mu)& = &\frac{\I}{2}\, B_\alpha +
\half  \sum_j \oint_{C_j} \frac{\de \varpi'}{2 \pi \I \varpi'} \,
\frac{\varpi' + \varpi}{\varpi' - \varpi}
\, \tilde{T}^{[+j]}_\alpha(\varpi')+  \ci{+}_\alpha  \log \varpi
+\cij{+}_\Lambda\(Y^\Lambda \varpi^{-1} + \bY^\Lambda \varpi\)
\, ,
\nonumber
\eea
where $\varpi$ is assumed to lie inside the contour $C_i$.
Finally, the contact potential is now given by
\bea
e^{\Phi\di{i}}&=&\left( \frac14 \sum_j\oint_{C_j}\frac{\de\varpi'}{2\pi\I \varpi'}
\(\varpi'^{-1} Y^{\Lambda}-\varpi' \bY^{\Lambda} \)
\tilde T^{[+j]}_\Lambda(\xi(\varpi'),\txi(\varpi'))
+\frac12\( \cij{+}_\Lambda A^\Lambda + \cij{+}_\alpha\)   \right)
\nonumber \\
&&\times\left(1+
\frac12 \sum_j \oint_{C_j} \frac{\de \varpi'}{2 \pi \I \varpi'} \,\frac{\varpi' + \varpi}{\varpi' - \varpi}
\, \p_{\ai{j}}\hHpij{+j}(\varpi') \right)
\, .
\label{eqchip}
\eea
The equations \eqref{txiqline}, \eqref{eqchip} provide sufficient information
to compute the deformed QK metric on $\cM$,
using the procedure outlined at the end of Subsection \ref{sect:2.1}.
As mentioned below \eqref{eqchi}, it may be convenient to trade the coordinate $\cR$
for the variable $\phi$, defined in the perturbed case by
 \be
\label{phipot}
\phi \equiv \Re\left[  {\Phi_{[+]}}(\varpi=0)   \right] \, .
\ee

Note that in general, the contact potentials $\Phi\di{i}$ are functions of $x^\mu$ and $\varpi$.
When $\hHpij{+j}$ is independent of $\ai{j}$, however, the analysis simplifies
considerably. As noted below \eqref{Sspe}, in this case
the contact potentials $\Phi\di{i}$ become all equal to a single real function on $\cM$,
which is nothing else but the function $\phi$ defined in \eqref{phipot}.
As we shall see, this situation
prevails for D-instanton corrections to the hypermultiplet branch, but  instanton
corrections with non-vanishing NS5-brane charge necessitate the
general formalism given here.

We end this executive summary of \cite{Alexandrov:2008nk}
with the following comment. All integration contours $C_j$
appearing in the formulae above
are closed since they surround open patches. It is however possible to generalize
\eqref{txiqline}, \eqref{eqchip} to open contours, provided the corresponding
transition functions $\hHij{+j}$ are finite at the endpoints.
This situation typically arises if one starts with a transition
function with a branch cut in the patch $\cU_j$, and shrinks the  contour around $\cU_j$
such that it surrounds the cut: the contribution reduces to the integral of the discontinuity
of the transition function (or appropriate combinations thereof) along the cut.
In this case the results \eqref{txiqline}, \eqref{eqchip}
acquire additional boundary contributions due to partial integrations, unless
$\hHij{+j}$ vanishes at the endpoints
of the open contour $C_j$.
For example, if $C_j$ is an open contour from $\varpi=0$ to $\varpi=\infty$,
the following term must be added to \eqref{eqchip},
\be
\frac{1}{8\pi \I}\(\hHij{+j}\vert_{\varpi=0}+\hHij{+j}\vert_{\varpi=\infty}\)\, .
\label{addcontr}
\ee
Thus, instead of assigning a set of open patches and transition functions,
a QK manifold can be characterized by providing a set of (closed or open) contours on $\CP$
and a set of associated functions $\hHij{+j}$. We will encounter such a description in the
discussion of the instanton corrected hypermultiplet moduli space in Section \ref{sec:Dinst}.

%------------------------------------------------------------
\subsection{Action of continuous isometries}
%------------------------------------------------------------

 We now discuss how isometries on ${\cal M}$ lift to
holomorphic isometries on its twistor space ${\cal Z}$.
This issue has been discussed in the literature before, see e.g.
\cite{MR872143,deWit:2001bk,Bergshoeff:2004nf}. Here we adapt it to our framework and make
some additional observations.

Suppose ${\cal M}$ admits a continuous isometry generated by a Killing vector field
$\kappa$. Generically, such an isometry rotates the quaternionic
two-forms ${\vec \omega}$ \eqref{su2curv} among each other,
\begin{equation}\label{isom-curv}
\cL_\kappa {\vec \omega} + {\vec r}\times {\vec \omega}=0\, ,
\end{equation}
where $\cL$ denotes the Lie derivative and $\vec{r}$ generates the rotation of the quaternionic two-forms.
The requirement that the two-forms  are
covariantly closed,
${\rm d}\vec\omega + \vec p \times\vec\omega = 0$,
determines the action of the Killing vector on  the $SU(2)$
connection
\begin{equation}\label{isom-conn}
\cL_\kappa \vec p = {\rm d}\vec r + \vec r \times \vec p\, .
\end{equation}
Under the action of the isometry $\kappa$, the second term in the
twistor space metric \eqref{Z-metric}
is invariant, but the projectivized connection $\mathcal{P}\equiv D\varpi-\de \varpi$
transforms due
to \eqref{isom-conn}. This can be remedied by combining the Killing action $\kappa$
on $\cM$ with a compensating $SU(2)$ rotation on the $\CP$ fiber, into
the vector field $\kappa_\cZ$ on $\cZ$
\begin{equation}
\label{defkz}
\kappa_\cZ = \kappa +
\left( r_+ -\I\, r_3 \,\varpi + r_- \varpi^2\right) \pa_{\varpi} +
\left( r_- +\I\, r_3 \,\bar{\varpi} +r_+ {\bar\varpi}^2\right) \pa_{\bar\varpi}\, .
\end{equation}
Under the Lie action of $\kappa_\cZ$, the canonical one-form $D\varpi$ transforms as
\begin{equation}
\label{varDz}
\cL_{\kappa_\cZ} \, D\varpi=(-\I\,r_3 +2\,r_- \varpi)\,D\varpi \, ,
\end{equation}
which ensures the invariance of the metric \eqref{Z-metric}. The vector $\kappa_\cZ$ is in fact
the real part of a holomorphic vector field $\kappa_h$ on $\cZ$, in accord with the fact
that any isometric action on $\cM$ can be lifted to an holomorphic action on
$\cZ$ \cite{MR872143,deWit:2001bk,Bergshoeff:2004nf}.

The vector $\vec r$ is
related to the vector-valued moment map $\vec \mu$ via \cite{MR872143}
\begin{equation}
\label{rmu}
\vec {\muh}  = \frac12 ( \vec r + \kappa \cdot \vec p )\, ,
\end{equation}
where $\vec p$ is the $SU(2)$ connection and the dot denotes the inner product.
The moment map provides a global holomorphic section of $H^0(\cZ,\cO(2))$,
\be
\label{defmu}
\muh\di{i} \equiv
e^{\Phi\di{i}} \left( \muh_+\, \varpi^{-1} -\I \muh_3 + \muh_- \varpi\right)\, .
\ee
To see that $\muh\di{i}$ is holomorphic, note that by virtue of \eqref{defkz}, \eqref{varDz} and
\eqref{rmu}, it equals the inner product of the Killing vector $\kappa_\cZ$ with the
 holomorphic one-form $\hCX\ui{i}$,
\be
\label{kzmu}
\kappa_\cZ \cdot \hCX\ui{i} = \muh\di{i}\, .
\ee
Since $\kappa_\cZ$ is the real part of a holomorphic vector field,
$\muh\di{i} $ is indeed a holomorphic function on $\hU_i$, hence defines
an element of $H^0(\cZ,\cO(2))$. Conversely, it is known that
any element of $H^0(\cZ,\cO(2))$ determines a
continuous isometry of $\cM$ \cite{MR664330}.

In fact, \eqref{kzmu} identifies $\muh\di{i}$ as the contact Hamiltonian
for the contact vector field $\kappa_\cZ$ \cite{MR2194671}, and
$\mu\di{i}^\cS\equiv \nui{i}^\alpha \muh\di{i}$ as (minus) the
complex moment map for the lift $\kappa_\cS$ of $\kappa$ to the Swann bundle. The Poisson
bracket associated to the complex symplectic structure on $\cZ_\cS$ descends to a
``contact Poisson'' bracket on $\cZ$, mapping
two local sections $(\muh_1, \muh_2)$ of $\cO(2m)\times \cO(2n)$ to a local
section of $\cO(2(m+n-1))$,
\be
\label{poisson}
\begin{split}
\{ \muh_1, \muh_2 \} &\equiv m\, \muh_1 \pa_\alpha \muh_2
+ \pa_\alpha \muh_1 \, \xi^\Lambda \pa_{\xi^\Lambda} \muh_2
+ \pa_{\xi^\Lambda}  \muh_1 \pa_{\txi_\Lambda}  \muh_2 \\
&\ -n\, \muh_1 \pa_\alpha \muh_2 -  \pa_\alpha \muh_2 \, \xi^\Lambda \pa_{\xi^\Lambda} \muh_1
- \pa_{\xi^\Lambda}  \muh_2 \pa_{\txi_\Lambda}  \muh_1\, .
\end{split}
\ee
For $m=n=1$, this defines a standard Poisson bracket on $H^0(\cZ,\cO(2))$,
such that, for two contact vector fields $\kappa_{1,2}$,
$\mu_{[\kappa_1,\kappa_2]}=\{ \mu_{\kappa_1},\mu_{\kappa_2}\}$. For $m=1,n=0$,
one obtains the  action of the Killing vector $\kappa_\cZ$ on the local complex coordinates,
\be
\label{mucontact}
\{ \muh, \xi^\Lambda \}= - \pa_{\txi_\Lambda}  \muh + \xi^\Lambda
\pa_{\alpha}  \muh\, ,
\qquad
\{ \muh, \txi_\Lambda \}= \pa_{\xi^\Lambda}  \muh \, .
\qquad
\{ \muh, \alpha \} = \muh  -  \xi^\Lambda  \pa_{\xi^\Lambda}  \muh \, .
\ee
This reproduces the vector field $(T_{[ij]}^\Lambda,\tilde{T}^{[ij]}_\Lambda,\tilde{T}^{[ij]}_\alpha)$
in \eqref{Tfct} for $\mu=\hHij{ij}+ \hHpij{ij}$.

Finally, inserting \eqref{varDz} into  \eqref{contact}, the holomorphic one-form transforms as
\begin{equation}
\label{isom-contact}
\cL_{\kappa_\cZ} \hCX\ui{i} =
\(\kappa_\cZ \cdot \Phi\di{i} + r_- \varpi - r_+ \varpi^{-1}\) \hCX\ui{i} =
\(\pa_{\ai{i}}  \muh\di{i}\)\hCX\ui{i}\, .
\end{equation}
This determines the variation of the contact and \kahler potentials to be
\be
\kappa_\cZ \cdot \Phi\di{i}  =  \pa_{\ai{i}}  \muh\di{i} - r_- \varpi + r_+ \varpi^{-1}\, ,
\qquad
\kappa_\cZ \cdot K_\cZ\ui{i}  = \Re\left( \pa_{\ai{i}}  \muh\di{i} \right)\, .
\label{transf_PhiK}
\ee

%------------------------------------------------------------
\subsection{Relation to Swann's construction}
%------------------------------------------------------------

As mentioned in the introduction, QK manifolds $\cM$ are in one-to-one correspondence
with \hk cones $\cS$ via the superconformal quotient and Swann's
constructions  \cite{MR1096180,deWit:2001dj}.
Such cones are completely
characterized by a single function, the \hk potential $\chi$, which
is a K\"ahler potential for the whole sphere of complex structures on $\cS$.
In \cite{Alexandrov:2008nk} it was shown that
$\chi$ is related to the contact potential via
\be
\chi=\frac{e^\phi}{4\rf}\, ,
\label{chiphi}
\ee
where $\phi$ is defined in \eqref{phipot}
and $\rf$ is a certain function invariant under the $SU(2)$ isometric action on $\cS$,
with weight one under dilations (see \cite{Alexandrov:2008nk} for more details).

The advantage of $\chi$ over $\phi$ is that it is invariant
under all isometries of $\cM$, while $\phi$ transforms non-trivially according to \eqref{transf_PhiK}.
This invariance was instrumental in the previous studies of instanton
corrections \cite{RoblesLlana:2006is,RoblesLlana:2007ae},
which are easily translated into our framework via \eqref{chiphi}.

%%%%%%%%%%%%%%%%%%%%%%%%%%%%%%%%%%%%%%%%%%%%%%%%%%%%%%%%%%%%%%%%%%%%%
\section{Perturbative hypermultiplet moduli spaces}
\label{sec_perturb}
%%%%%%%%%%%%%%%%%%%%%%%%%%%%%%%%%%%%%%%%%%%%%%%%%%%%%%%%%%%%%%%%%%%%%

In this section we recall some known results on the perturbative
hypermultiplet moduli space in Type IIA
and IIB string theories compactified on a CY threefold,
and phrase them in the language of twistors and
complex contact geometry.

%------------------------------------------------------------
\subsection{Type IIA compactified on a CY threefold $X$\label{seciia}}
%------------------------------------------------------------

The hypermultiplet moduli space $\cM_{\rm HM}^A$ in Type IIA string theory
compactified on a CY
threefold $X$ is a QK manifold of real
dimension $d=4(h_{2,1}(X)+1)$ \cite{Cecotti:1988qn,Ferrara:1989ik,Bodner:1990zm}.
It describes the dynamics of the complex structure moduli
$X^\Lambda=\int_{\gamma^\Lambda} \Omega$,
$F_\Lambda=\int_{\gamma_\Lambda} \Omega$, the RR scalars
\be
\label{rriia}
\zeta^\Lambda=\int_{\gamma^\Lambda} A^{(3)} \ ,\qquad
\tzeta_\Lambda=\int_{\gamma_\Lambda} A^{(3)}\ ,
\ee
the four-dimensional dilaton
$e^{\phi}=1/g_{(4)}^2$ and the Neveu-Schwarz (NS) axion $\sigma$, dual to the
Neveu-Schwarz two-form $B$ in four dimensions. Here $\gamma^\Lambda$
and $\gamma_\Lambda$ form a symplectic basis
of A and B cycles in $H_3(X,\IZ)$, with
intersection product $\langle \gamma^\Lambda, \gamma_\Sigma\rangle=
\delta^\Lambda_\Sigma$.

To all orders in perturbation theory, the metric on $\cM_{\rm HM}^A$ admits
a $2d+1$-dimensional Heisenberg group of tri-holomorphic isometries, corresponding
to translations along the RR potentials $(\zeta^\Lambda, \tzeta_\Lambda)$ and the NS axion $\sigma$.
Thus, it falls into the class of toric QK geometries discussed in Section \ref{sectoric}.
These continuous isometries are in general broken to a discrete subgroup by
instanton corrections. At tree level, the geometry of $\cM_{\rm HM}^A$ is obtained from
the moduli space $\cM_{\rm cs}$ of complex structure  deformations of $X$
(which would correspond to the vector multiplet moduli space of Type IIB string theory
compactified on the {\it same} CY $X$)
via the ``$c$-map'' construction \cite{Cecotti:1988qn,Ferrara:1989ik}.
$\cM_{\rm cs}$ is completely characterized by the prepotential $F(X^\Lambda)$,
a homogeneous function of degree 2 of the A-type periods $X^\Lambda$, such that
the B-type periods are given by
$F_\Lambda=\pa F/\pa X^\Lambda$.  $X^\Lambda$ provide
a set of homogeneous coordinates on $\cM_{\rm cs}$, and (away from the
vanishing locus of $X^0$) may be traded for
the inhomogeneous coordinates $z^a=X^a/X^0$. At one-loop, the $c$-map
metric on $\cM$ receives a correction proportional to the
Euler class $\chi_X=2(h^{1,1}(X)-h^{2,1}(X))$.

The twistor space $\cZ$ of the QK manifold $\cM_{\rm HM}^A$ admits the following simple
description. $\cZ$ can be covered by two patches $\hU_+$, $\hU_-$
which project to open disks centered around $\varpi=0$ and $\varpi=\infty$ on $\CP$,
and a third patch $\hU_0$ which projects to the rest of $\CP$. The transition functions
between complex Darboux coordinates on each patch are given by \cite{Alexandrov:2008nk}
\be
\label{symp-cmap}
\hHij{0+}= -\frac{\I}{2}\, F(\xi^\Lambda)\, ,
\qquad
\hHij{0-}=-\frac{\I}{2}\,\bF(\xi^\Lambda)\, ,
\qquad
\ci{+}_\alpha = \frac{\chi_X}{96\pi} \, ,
\ee
with the other anomalous dimensions  $\ci{\pm}_\Lambda=0$. The non-vanishing anomalous
dimension $\ci{\pm}_\alpha$ incorporates the effect of the one-loop correction.
Based on the
string theory amplitudes \cite{Antoniadis:1997eg,Gunther:1998sc},
the QK metric obtained from \eqref{symp-cmap} was calculated
in \cite{Robles-Llana:2006ez,Alexandrov:2007ec}.
It is believed to be the correct metric on $\cM$ to all orders in perturbation
theory \cite{Gunther:1998sc,Robles-Llana:2006ez,Alexandrov:2008nk}.
At this perturbative level,
the coordinates $Y^a,A^\Lambda, B_I$ introduced on general grounds
in Section 2 are related to the  Type IIA variables via
\be
\zeta^\Lambda=A^\Lambda\, ,
\quad
\tzeta_\Lambda=B_{\Lambda}+A^\Sigma \Re F_{\Lambda\Sigma}(z)\, ,
\quad
\sigma =- 2 B_\alpha  - A^\Lambda B_\Lambda\, ,
\quad
Y^a = \cR\, z^a \, ,
\label{relABzeta}
\ee
where $\cR$ may be expressed in terms of the contact
potential by means of \eqref{eqchi},
 \be
 \label{phipertA}
e^{\Phi_{\rm pert}}
= \frac{\cR^2}{4}\,K(z,\bz)+ \frac{\chi_X}{192\pi}
\ee
with $K(z,\bz)\equiv-2 \Im(\bz^\Lambda F_\Lambda)$. The contact potential $\Phi_{\rm pert}$
is in turn identified with the 4D dilaton $\phi$.
Denoting
\be
\label{defrho}
\rho_\Lambda\equiv  -2\I \txii{0}_\Lambda\, ,
\qquad
\tilde\alpha\equiv 4\I  \ai{0} + 2 \I \txii{0}_\Lambda \xi^\Lambda\, ,
\ee
the contact twistor lines in the patch $\hU_0$ are given by \cite{Neitzke:2007ke,Alexandrov:2008nk}
\be
\label{gentwi}
\begin{array}{rcl}
\xi^\Lambda &=& \zeta^\Lambda + \cR
\left( \varpi^{-1} z^{\Lambda} -\varpi \,\bz^{\Lambda}  \right)\, ,
\\
\rho_\Lambda &=& \tzeta_\Lambda + \cR
\left( \varpi^{-1} F_\Lambda(z)-\varpi \,\bF_\Lambda(\bz) \right)\, ,
\\
\tilde\alpha&=& \sigma + \cR
\left(\varpi^{-1} W(z)-\varpi \,\bar W(\bz) \right) +\frac{\I\chi_X}{24\pi} \,\log \varpi \, ,
\end{array}
\ee
where
\be
\label{defW}
W(z) \equiv  F_\Lambda(z) \zeta^\Lambda - z^\Lambda \tzeta_\Lambda\, .
\ee

Electric-magnetic duality acts on $\cZ$ by complex
contact transformations
\be
\label{emag}
\begin{pmatrix}
\xi^\Lambda\\
\rho_\Lambda
\end{pmatrix} \mapsto
\begin{pmatrix} \CA & \CB \\ \CC & \CD \end{pmatrix}
\begin{pmatrix}
\xi^\Lambda\\
\rho_\Lambda
\end{pmatrix}\, ,
\qquad
\tilde\alpha \mapsto \tilde\alpha\, ,
\ee
where $\scriptsize\begin{pmatrix} \CA & \CB \\ \CC & \CD \end{pmatrix}$ is a $Sp(2h_{1,2}(X),\IZ)$
matrix whose block matrices satisfy
\be
\CA^{\rm T} \CC - \CC^{\rm T} \CA = \CB^{\rm T} \CD - \CD^{\rm T} \CB = 0 \, ,
\qquad
\CA^{\rm T} \CD - \CC^{\rm T} \CB =  \bf{1} \, .
\ee
This action is in general not an isometry of $\cM$, since the moment map
associated to an infinitesimal action \eqref{emag}
with $\CB=\CB^{\rm T},\ \CC=\CC^{\rm T},\ \CA^{\rm T}+\CD=0$, given by
\be
\muh = - \txii{0}_\Lambda \CA^{\Lambda}_{\,\,\Sigma} \xi^\Sigma
+ \I \,\txii{0}_\Lambda \, \CB^{\Lambda\Sigma} \, \txii{0}_\Sigma
+\frac{\I}{4}\,  \xi^\Lambda \, \CC_{\Lambda\Sigma} \, \xi^\Sigma\, ,
\ee
is in general {\it not} a global $\cO(2)$ section.\footnote{For special choices of $\kappa_{abc}$,
related to Jordan algebras of degree 3, a subgroup of $Sp(2h_{1,2}(X),\IZ)$
may however act isometrically, see e.g. the $SL(2,\IR)$
generators $Y_+,Y_0,Y_-$ in
eq.\ (3.50) of \cite{Gunaydin:2007qq} for
the special case $\cM=G_{2(2)}/SO(4)$.}

%------------------------------------------------------------
\subsection{Type IIB compactified on a CY threefold $Y$ \label{seciib}}
%------------------------------------------------------------

The hypermultiplet moduli space $\cM=\cM_{\rm HM}^B$ in
Type IIB string theory compactified on a CY
threefold $Y$ is a QK manifold of
real dimension $d=4(h^{1,1}(Y)+1)$ \cite{Cecotti:1988qn,Bodner:1989cg,Ferrara:1989ik,
Gunther:1998sc}.
It describes the dynamics of the
K\"ahler moduli $z^a\equiv b^a + \I t^a=\int_{\gamma^a} \mathcal{J}$,
the  RR scalars\footnote{The $B$-dependent corrections ensure that
these fields have simple transformation properties under S-duality, see eq.\ \eqref{SL2Z} below.
The correction to $c_a$ appears in \cite{Louis:2002ny}, footnote 14, the correction
to $c_0$ seems to be novel.}
\be
\label{RRiib}
\begin{split}
c^0=A^{(0)}\, ,
& \qquad
c^a=\int_{\gamma^a} A^{(2)}\, ,\qquad
c_a=- \int_{\gamma_a} (A^{(4)} - \frac{1}{2}\, B \wedge A^{(2)})\, ,
\\
& c_0=-\int_{Y} (A^{(6)} - B \wedge  A^{(4)} + \frac{1}{3}\, B \wedge B \wedge A^{(2)} )\, ,
\end{split}
\ee
the four-dimensional dilaton $\phi$ and the NS axion $\psi$, dual to the NS 2-form
$B$ in four dimensions.
Here $\mathcal{J}\equiv B+\I\, J=z^a \omega_a$ is the
complexified \kahler form on $Y$. Furthermore,  $\gamma^a$,
$a=1,\dots, h^{1,1}(Y)$, denote a basis of 2-cycles (Poincar\'e dual to 4-forms $\omega^a$), and
$\gamma_a$ a basis of
4-cycles (Poincar\'e dual to 2-forms $\omega_a$), such that
\be\omega_a \wedge\omega_b = \kappa_{abc}\omega^c\, ,
\qquad
\omega_a \wedge \omega^b = \delta_a^b \omega_Y\, ,
\qquad
 \int_{\gamma^a}\omega_b= \int_{\gamma_b}\omega^a= \delta^a_b\, ,
\ee
where $\omega_Y$ is the volume form, normalized to $\int_Y \omega_Y =1$,
and $ \kappa_{abc}=\int_Y \omega_a \omega_b \omega_c
=\langle \gamma_a, \gamma_b, \gamma_c\rangle$
is the triple intersection product in $H_4(Y,\IZ)$.
In the large volume limit, the 4D dilaton $\phi$ is related to the 10D
string coupling $g_s$ via $e^{\phi}=V(t^a)/g_s^2$, where
$V(t^a)=\frac{1}{6}\int_Y J\wedge J\wedge J = \frac16 \kappa_{abc}t^a t^b t^c$
is the volume of $Y$ in string units. The  ten-dimensional
coupling $\tau_2\equiv 1/g_s$ and the RR axion $\tau_1\equiv c^0$
can be combined into the ten-dimensional axio-dilaton field $\tau = \tau_1+\I \tau_2$.

As in Type IIA string theory, to all orders in perturbation theory, the metric on $\cM_{\rm HM}^B$ admits
a $2d+1$-dimensional Heisenberg group of  isometries, corresponding
to translations along the RR potentials $(c^\Lambda, c_\Lambda)$ and the NS axion $\psi$.
At tree level, it is obtained from the moduli space $\cM_{\rm ks}$ of complexified K\"ahler deformations
(which would correspond to the vector multiplet moduli space in Type IIA string theory
compactified on the {\it same} CY $Y$) via the $c$-map.
$\cM_{\rm ks}$ is again characterized by the prepotential $F(X^\Lambda)$, which now
receives world-sheet instanton corrections. The prepotential has the standard large volume
expansion (in the conventions of \cite{Saueressig:2007dr}, up to a sign change in
$\kappa_{abc}$)),
\be
\label{lve}
F(X^\Lambda)=-\kappa_{abc} \,\frac{X^a X^b X^c}{6 X^0} + \chi_Y\,
\frac{\zeta(3)(X^0)^2}{2(2\pi\I)^3}
-\frac{(X^0)^2}{(2\pi\I)^3}{\sum_{k_a\gamma^a\in H_2^+(Y)}} n_{k_a}^{(0)}\, \Li_3\left(
e^{2\pi \I  k_a X^a/X^0}\right)\, ,
\ee
where $k_a$ runs over effective homology classes
(i.e. $k_a\geq 0$ for all $a$, not all of them vanishing simultaneously),
$n_{k_a}^{(0)}$ is the genus zero BPS invariant in the homology
class $k_a \gamma^a \in H_2^+(Y,\IZ)$,
$\Li_s(x)=\sum_{m=1}^\infty m^{-s}x^m$ is the polylogarithm function,
and $\chi_Y$ is the Euler number of $Y$. Note that the last two terms
in \eqref{lve} may be combined by including the zero class
$k_a=0$ in the sum and setting $n_0^{(0)}=-\chi_Y/2$.

At one-loop, the $c$-map metric on $\cM$ receives a correction proportional to the Euler
class $\chi_Y$ of $Y$. The twistor space $\cZ$ is described by the same transition
functions  \eqref{symp-cmap}  and contact potential  in \eqref{phipertA}, with
$\chi_X$ replaced by $-\chi_Y$. Using the large volume expansion \eqref{lve}
and identifying $\cR=\tau_2/2$ (as will become clear in \eqref{physmap} below), the contact potential
can be further expressed as
\be
\label{phipertB}
e^{\Phi_{\rm pert}}= \frac{\tau_2^2}{2} \,V(t^a)-\frac{\chi_Y\zeta(3)}{8(2\pi)^3}\,\tau_2^2
+ e^{\Phi_{\rm ws}}- \frac{\chi_Y}{192\pi}\, ,
\ee
where
\be
\label{phiws}
e^{\Phi_{\rm ws}} =\frac{\tau_2^2}{4(2\pi)^3}\sum_{k_a\gamma^a\in H_2^+(Y)} n_{k_a}^{(0)}
\Re\left[ \Li_3 \left( e^{2\pi \I k_a z^a} \right) + 2\pi k_a t^a\,
\Li_2 \left( e^{2\pi \I k_a z^a} \right)  \right]
\ee
is the world-sheet instanton  contribution.
In the large volume limit, $\Phi_{\rm pert}$ coincides with the 4D dilaton $\phi$,
and may in fact be taken as the definition of the 4D dilaton in the quantum regime.

When $X$ and $Y$ are related by mirror symmetry (which requires $\chi_X=-\chi_Y$),
the hypermultiplet moduli
spaces $\cM_{\rm HM}^A$ and $\cM_{\rm HM}^B$ must be identical, with the
\kahler moduli $z^a=b^a+\I t^a$ of $Y$ being identified with the complex structure
moduli of $Y$. The relation between the Type IIA variables
$(\cR,Y^a,\zeta^\Lambda,\tzeta_\Lambda,\sigma)$
and the Type IIB variables $(\tau, z^a, c^a, c_a, c_0, \psi)$ will be obtained in the next subsection
from S-duality.

%------------------------------------------------------------------------
\subsection{S-duality and mirror map}
%------------------------------------------------------------------------

At the classical level, i.e., at tree-level and leading order in the $\alpha'$ expansion, Type IIB
supergravity in ten dimensions is invariant under a continuous $SL(2,\IR)$ symmetry.
After compactification on $Y$ and in the large volume limit,  the metric on the hypermultiplet
moduli space $\cM\equiv \cM_{\rm HM}^B$ admits an isometry group $SL(2,\IR)$,
acting as
\be\label{SL2Z}
\begin{split}
&\quad \tau \mapsto \frac{a \tau +b}{c \tau + d} \, ,
\qquad
t^a \mapsto t^a |c\tau+d| \, ,
\qquad
c_a\mapsto c_a\, ,
\\
&
\begin{pmatrix} c^a \\ b^a \end{pmatrix} \mapsto
\begin{pmatrix} a & b \\ c & d  \end{pmatrix}
\begin{pmatrix} c^a \\ b^a \end{pmatrix}\, ,
\qquad
\begin{pmatrix} c_0 \\ \psi \end{pmatrix} \mapsto
\begin{pmatrix} d & -c \\ -b & a  \end{pmatrix}
\begin{pmatrix} c_0 \\ \psi \end{pmatrix}
\end{split}
\ee
with $ad-bc=1$. While the existence of this isometric action was
established in \cite{Gunther:1998sc,Bohm:1999uk}, it is instructive to derive it again
by twistorial methods.

For this purpose, it suffices, as explained below \eqref{defmu},
to construct global $\cO(2)$ sections whose
Poisson brackets satisfy the $SL(2,\IR)$ algebra.
The following three quantities\footnote{These formulae agree
with the moment maps computed in \cite{Gunaydin:2007qq} for
the special case $\cM=G_{2(2)}/SO(4)$; compare \eqref{SL2inf0} with
the generators $E_{p^0}, H+2Y_0,F_{p^0}$ in
eq.\ (3.50) of \cite{Gunaydin:2007qq}.}
\be
\label{SL2infp}
\muh^+ = -\txii{+}_0 - \frac{\I}{12(\xi^0)^2}\, \kappa_{abc}\, \xi^a\xi^b\xi^c\ ,\quad
\muh^0 = \ai{+}- \xi^0 \txii{+}_0\, ,
\qquad
\muh^-  = \ai{+} \xi^0
\ee
satisfy these requirements. Indeed, they are manifestly regular at $\varpi=0$,
(except for the $1/\varpi$ pole that we allow for global $\cO(2)$ sections,
as in e.g. \eqref{gxi}).
The apparent singularity at the zeros of $\xi^0$ can be removed by
rewriting them in the  patch $\hU_0$, using the classical limit of the transition functions \eqref{symp-cmap},
\be
\label{SL2inf0}
\muh^+ = -\txii{0}_0 \, ,
\qquad
\muh^0 = \ai{0}- \xi^0 \txii{0}_0\, ,
\qquad
\muh^-  = \ai{0} \xi^0 - \frac{\I}{12} \kappa_{abc}\, \xi^a\xi^b\xi^c\, .
\ee
The regularity at $\varpi=\infty$ is of course guaranteed by the reality condition.
Exponentiating the infinitesimal action generated by
the contact Hamiltonians \eqref{SL2inf0}, we arrive at
the $SL(2,\IR)$ action on the contact twistor lines in the patch $\hU_0$,
\be
\label{SL2Zxi}
\begin{split}
\xi^0 &\mapsto \frac{a \xi^0 +b}{c \xi^0 + d} \, , \qquad
\xi^a \mapsto \frac{\xi^a}{c\xi^0+d} \, , \qquad
\txi_a \mapsto \txi_a +  \frac{\I\, c}{4(c \xi^0+d)} \kappa_{abc} \xi^b \xi^c\, ,
\\
\begin{pmatrix} \txi_0 \\ \alpha \end{pmatrix} &\mapsto
\begin{pmatrix} d & -c \\ -b & a  \end{pmatrix}
\begin{pmatrix} \txi_0 \\ \alpha \end{pmatrix}
+ \frac{\I}{12} \kappa_{abc} \xi^a\xi^b\xi^c
\begin{pmatrix}
c^2/(c \xi^0+d)\\
-[ c^2 (a\xi^0 + b)+2 c] / (c \xi^0+d)^2
\end{pmatrix}\, .
\end{split}
\ee
The action on $\xi^\Lambda$ agrees with  the standard
linear action on the complex coordinates $\nu^I$ on the Swann bundle
\cite{Berkovits:1998jh,RoblesLlana:2006is} after projectivizing.
Under the action \eqref{SL2Zxi},  the complex
contact one-form transforms by an overall holomorphic factor
$\hCX\ui{i}\to \hCX\ui{i}/(c\xi^0+d)$, leaving the complex
contact structure invariant.

The holomorphic contact action \eqref{SL2Zxi} on $\cZ$ descends to an isometric action on
$\cM$, and a $SU(2)$ rotation on the fiber. It may be checked that \eqref{SL2Zxi} agrees with
the standard action \eqref{SL2Z}, provided one identifies\footnote{The coefficient of the first term
in $\tzeta_0$ and $\sigma$ cannot be determined from $SL(2,\IR)$ invariance alone,
as it can be changed by a field redefinition
shifting $(\txi_0,\alpha)$ by a term proportional to the doublet
$(c_0,\psi)$. It is fixed however by requiring the consistency of the D-brane actions
\eqref{Sin} and \eqref{Sbbrane} under mirror symmetry.}
\be
\label{physmap}
\begin{split}
\cR& =\frac12\,\tau_2\, , \qquad
Y^a=\frac12\, \tau_2 \, z^a\, ,\qquad
\zeta^0=\tau_1\, ,
\qquad
\zeta^a = - (c^a - \tau_1 b^a)\, ,
\\
\tzeta_a &=  c_a+ \frac{1}{2}\, \kappa_{abc} \,b^b (c^c - \tau_1 b^c)\, ,
\qquad
\tzeta_0 =\, c_0-\frac{1}{6}\, \kappa_{abc} \,b^a b^b (c^c-\tau_1 b^c)\, ,
\\
\sigma &= -2 (\psi+\frac12  \tau_1 c_0) + c_a (c^a - \tau_1 b^a)
-\frac{1}{6}\,\kappa_{abc} \, b^a c^b (c^c - \tau_1 b^c)\, .
\end{split}
\ee
These relations, valid in the classical limit, provide the ``generalized mirror map"
between the Type IIA variables $(\cR,Y^a,\zeta^\Lambda,\tzeta_\Lambda,\sigma)$ and
the Type IIB variables $(\tau, b^a, t^a, c^a,$ $ c_a,c_0,\psi)$. They agree
with the identification found by dimensional reduction of the Type IIB supergravity
Lagrangian on $Y$ in  \cite{Gunther:1998sc,Bohm:1999uk}.\footnote{
Note that this identification was derived independently on the
vector multiplet side in the one-modulus case
in \cite{Berkooz:2008rj}, eq. (3.13-14). The identifications are $(V,\rho_2,\rho_1,\mu_1,\mu_2,\nu,
\tilde\mu_1,\tilde\mu_2)_{\rm  \cite{Berkooz:2008rj}}=
(\tau_2 (t^a)^2, e^{\phi/2}/\tau_2^{3/2},-\tau_1,\sqrt3 c^a, \sqrt3 b^a, c_a/\sqrt3,
-\psi/\sqrt2, c_0/\sqrt2)_{\rm Here}$.
}

Expressing $\varpi$ in terms of $\xi^0$ and using the first equation in \eqref{SL2Zxi},
we obtain the action of $SL(2,\IR)$ on the $\CP$ fiber,
\be
\varpi \mapsto  \frac{c \tau_2 + \varpi  (c \tau_1 + d) +
\varpi |c \tau + d| }{(c \tau_1 + d) + |c \tau + d| - \varpi c \tau_2}\, .
\ee
Moreover, the contact potential transforms as
\be
\label{SL2phi}
e^\Phi \mapsto e^\Phi/ |c\tau+d|\, ,
\ee
which ensures that the \kahler potential varies by a \kahler transformation,
\be
K_\cZ\mapsto  K_\cZ - \log(|c\xi^0+d| )\, .
\ee

In the presence of worldsheet instantons or after including the one-loop correction,
the continuous isometries associated to $\mu^-$ and $\mu^0$
are broken since their
purported moment maps are no longer regular at $\varpi=0$. As shown in
\cite{RoblesLlana:2006is,RoblesLlana:2007ae} and reviewed in Section \ref{subsec_2B}
below, it is possible to restore the invariance under a discrete subgroup
$SL(2,\IZ)$ by incorporating D-instanton corrections.

%%%%%%%%%%%%%%%%%%%%%%%%%%%%%%%%%%%%%%%%%%%%%%%%%%%%%%%%%%%%%%%%%%%%%%%
\section{D-instanton corrections in Type II compactifications}
\label{sec:Dinst}
%%%%%%%%%%%%%%%%%%%%%%%%%%%%%%%%%%%%%%%%%%%%%%%%%%%%%%%%%%%%%%%%%%%%%%%

In this section, we determine the form of all D-instanton corrections to the hypermultiplet
metric, as linear perturbations around the perturbative QK metric. We start by
reviewing and extending the results obtained  in \cite{RoblesLlana:2006is,RoblesLlana:2007ae}
for the contribution of  D1-D$(-1)$-instantons and  A-type D2-brane instantons
in Calabi-Yau compactifications of Type IIB and Type IIA strings, respectively.
We then generalize these results to all D-instantons, using electric-magnetic duality
and mirror symmetry. In Subsection \ref{secrig} we extend our considerations beyond
linear order. 

%------------------------------------------------------------------------
\subsection{S-duality and A-type D-instanton corrections}
%------------------------------------------------------------------------
\label{subsec_2B}

While \eqref{phipertB} is believed to be the full perturbative result, it cannot be exact,
since it is not consistent with  $SL(2,\IZ)$ duality of ten-dimensional Type IIB string theory.
Indeed, the subleading terms in  \eqref{phipertB} spoil the transformation rule
\eqref{SL2phi}, and the \kahler potential no longer transforms by a \kahler transformation.

As explained in  \cite{RoblesLlana:2006is}, the invariance under the discrete
subgroup $SL(2,\IZ)$ can be
restored by summing over images, using similar techniques as the ones used for
$R^4$ couplings in toroidal compactifications
\cite{Green:1997tv,Green:1997as,Kiritsis:1997em,Obers:1999um}.
The result is expressed in terms of a generalized Eisenstein series,
\be
e^{\Phi_{\rm inv}} = \frac{\tau_2^2}{2} \,V(t^a)
+\frac{\sqrt{\tau_2}}{8(2\pi)^3}
\!\!\sum_{k_a\gamma^a\in H_2^+(Y)\cup\{0\}}\!\!\!\!\!\,
 n_{k_a}^{(0)}\,
{\sum\limits_{m,n}}'\frac{\tau_2^{3/2}}{|m\tau+n|^3}\(1+2\pi |m\tau+n|k_a t^a\)e^{-S_{m,n, k_a}}\, ,
\label{phiinv}
\ee
where
\be
S_{m,n,k_a} =2\pi k_a | m \tau+n |\, t^a-2\pi \I k_a (m c^a +n  b^a)
\ee
and the primed sum runs over pairs of integers $(m,n)\neq (0,0)$.
For $k_a= 0$ the sum encodes the perturbative
contributions together with the D$(-1)$-instanton corrections, while for $k_a\gamma^a\in H_2^+(Y)$,
the exponent $S_{m,n,k_a}$ is the classical action of a $(p,q)$-string
(or rather $(m,n)$-string) wrapped on the 2-cycle $k_a \gamma^a$.

Although the representation \eqref{phiinv} makes S-duality invariance manifest, it
cannot be directly interpreted as an instanton sum.  To expose the instantons, it is advisable
to perform a Poisson resummation on the integer $n$ \cite{Green:1997tv}.
Denoting the dual integer by $k_0$,
one obtains~\cite{RoblesLlana:2007ae}
 \be
\begin{split}
e^{\Phi_{\rm inv}} =& \frac{\tau_2^2}{2} \,V\left(t^a\right)
- \frac{\sqrt{\tau_2}}{8(2\pi)^3}\,\chi_Y
\left[ \zeta(3)\,\tau_2^{3/2}+\frac{\pi^2}{3}\,\tau_2^{-1/2}
\right]+ e^{\Phi_{\rm ws}}
\\
& +\frac{\tau_2}{8\pi^2}{\sum\limits_{k_\Lambda}}'
n_{k_a}^{(0)}
\sum_{m=1}^\infty
\frac{|k_\Lambda z^\Lambda|}{m}\, \cos\left(2\pi m \,k_\Lambda  \zeta^\Lambda\right)
K_1\left(2\pi m \, | k_\Lambda z^\Lambda|\tau_2\right)\, ,
\end{split}
\label{phiinst}
\ee
where we denoted
$k_\Lambda=(k_0,k_a)$. In this equation, the sum runs over $k_0\in\IZ$,
$k_a\gamma^a\in H_2^+(Y,\IZ)$ excluding the value $(k_0,k_a)=0$  (as indicated by
the prime),
and $e^{\Phi_{\rm ws}}$ is the world-sheet instanton contribution
\eqref{phiws}.
The  term in square brackets in \eqref{phiinst}  combines  two perturbative contributions:
the first is perturbative in the $\alpha'$ expansion
and corresponds to the second term in \eqref{phipertB},
whereas the second is the one-loop contribution corresponding to
the last term in \eqref{phipertB}.
In contrast, the second line in \eqref{phiinst} has a non-perturbative origin: it
describes the contributions of ``bound states'' of $m$ Euclidean D1-strings wrapping
rational curves (counted by the Gopakumar-Vafa invariant $n_{k_a}^{(0)}$)
in the homology class $k_a \gamma^a$ and $m\, k_0$ D$(-1)$-instantons, with classical action
\be
\label{clasact}
S_{\rm cl}=  2\pi m \tau_2 \,| k_\Lambda z^\Lambda|  + 2\pi \I m \,k_\Lambda  \zeta^\Lambda\, .
\ee
In the subsector $k_a=0$, the sum reduces to D$(-1)$-instanton contributions,
analogous to the ones appearing in $R^4$ couplings in ten dimensions \cite{Green:1997tv}.

Using the mirror map \eqref{physmap}, the same result \eqref{phiinst}
can be interpreted from the point of view of Type IIA string theory compactified
on $X$  \cite{RoblesLlana:2007ae}: the classical instanton action \eqref{clasact}
corresponds to a bound state of $m$ D2-branes wrapping the
A-cycle $k_\Lambda \gamma^\Lambda$ in $H_3(X,\IZ)$. By mirror symmetry,
the BPS invariant  $n_{k_a}^{(0)}$ of $Y$ should count  the number of special
Lagrangian 3-cycles homologous to $k_\Lambda \gamma^\Lambda$ in $H_3(X,\IZ)$;
in particular, this number should be independent of $k_0$.
Of course, on the Type IIA side the restriction to A-cycles
is artificial, and  will be relaxed in the next subsection.

Leaving a more detailed discussion of the instanton effects  to Section \ref{secDinst},
we now discuss how  the contact potential \eqref{phiinst} may be understood
from the twistor approach.\footnote{A contour integral presentation of \eqref{phiinv} of the form
\eqref{eqchi} was given in \cite{RoblesLlana:2007ae}, but its twistorial
interpretation is obscured by issues of convergence.} For this purpose,
let us define
\be
G_{\rm A}(\xi)=
\frac{1}{(2\pi)^2}
{\sum\limits_{(k_\Lambda)_+}} \hn_{k_\Lambda} \Li_2\left(
e^{-2\pi \I k_\Lambda \xi^\Lambda} \right)\, ,
\label{prepGP}
\ee
where the sum runs over the set (here $H_2^-(Y)=-H_2^+(Y)$)
\be
(k_\Lambda)_+ \equiv \{
k_0\in \IZ, \quad k_a\gamma^a \in H_2^+(Y) \cup H_2^-(Y) \cup \{0\} \ ,\quad
\Re \(k_\Lambda z^\Lambda\)>0\} \ .
\label{rangep}
\ee
This sum \eqref{prepGP} depends on the
value of the coordinates $z^\Lambda$ on $\cM$ through the last
condition in \eqref{rangep}, and through the coefficients $ \hn_{k_\Lambda}$;
the latter are locally constant away from
the ``lines of marginal stability" (LMS) where $\Re(k_\Lambda z^\Lambda)$ vanishes
for a certain vector $k_\Lambda$, but may change across the LMS. In order
to reproduce \eqref{phiinst} in the region connected to the infinite volume limit,
we require
\be
\hn_{(k_0,k_a)}= n_{k_a}^{(0)} \quad\mbox{\rm for}\quad k_a\gamma^a\ne 0\, , \qquad
\hn_{(k_0,0)}=2n_{0}^{(0)}=-\chi_Y\, .
\ee
Then, to the three-patch covering and transition functions \eqref{symp-cmap} describing the
perturbative moduli space, we add two additional transition functions
\be
\label{symp2-inst}
\hHij{0\ell_+}=-\frac{\I}{2}\,G_{\rm A}(\xi)\, ,
\qquad
\hHij{0\ell_-}=-\frac{\I}{2}\,\bG_{\rm A}(\xi)\, ,
\ee
which are associated
with {\it open} contours extending from $\varpi=0$ to $\varpi=\infty$
along the semi-infinite imaginary axes $\ell_\pm\equiv  \I\IR^{\pm}$.
This construction requires the extension of our formalism to open contours,
as discussed at the end of Section \ref{subsec_lindef}. Using
the integral representation
\be
\frac{1}{4}\int_0^\infty \frac{\de t}{t}\(\alpha t +\frac{\beta}{t}\)
e^{-\frac12\(\alpha t +\frac{\beta}{t}\)}
=\sqrt{\alpha\beta}\,K_1\(\sqrt{\alpha\beta}\)
\ee
valid for $\Re \alpha>0, \ \Re \beta>0$, it is then easy to see that
\eqref{eqchi} precisely reproduces the contact potential \eqref{phiinst}.
A more complete analysis of the structure of the twistor space will be given after we
incorporate the B-type D-instantons.

Note that the one-loop string correction,
which follows here from the non-vanishing anomalous dimension $\ci{\pm}_\alpha$,
can be obtained alternatively by adding the term $k_\Lambda=0$ to the sum in \eqref{prepGP}.
Since this constant term does not vanish at the ends of the contour,
as mentioned in the end of Section \ref{subsec_lindef},
the contact potential \eqref{eqchi} receives additional contributions
given in \eqref{addcontr} where $j=\ell_\pm$.
Taking $\hn_{(0,0)}=-\chi_Y/2$, it is easy to check that
these contributions reproduce the one-loop term in \eqref{phiinst}.

%------------------------------------------------------------------------
\subsection{Covariantizing under electric-magnetic duality}
\label{linins}
%------------------------------------------------------------------------

In general, instanton corrections break all continuous isometries of $\cM$
to a discrete subgroup. Thus, hypermultiplets can no longer be dualized to
tensor multiplets, and the projective superspace description in terms of the
$\cO(2)$ multiplets breaks down. However, as
explained in Section \ref{sec_projdescr}, linear perturbations of toric QK
manifolds can still be described by a set of generating functions
$\hHpij{ij}$, which now depend on all complex coordinates
$\xi^\Lambda,\ \txi_\Lambda$ and $\alpha$ on $\cZ$.

In the case of D-instantons, i.e. Euclidean D-branes
wrapping arbitrary cycles in $H_3(X,\IZ)$ (in Type IIA string theory)
or $H_{\rm even}(Y,\IZ) $ (in Type IIB string theory),  the translational isometry along
the NS-axion is preserved, and one
should therefore restrict to perturbations which are independent of $\alpha^{[j]}$.
As explained in Section \ref{subsec_lindef}, this considerably simplifies
the analysis, since e.g. the geometry of $\cZ$ can be described by
a single contact potential, as in the
unperturbed case. Moreover, while we may in principle perturb of
the A-instanton corrected toric geometry, we choose to treat
{\it all} instantons as perturbations around the perturbative
geometry described by the transition functions
given in \eqref{symp-cmap}.

Using the action of electric-magnetic duality described in Section \ref{seciia},
the function \eqref{prepGP} describing the D-instanton part
may be covariantized into
\be
G_{\rm A/B}(\xi^\Lambda,\rho_\Lambda)=
\frac{1}{(2\pi)^2}
\ {\sum\limits_{\scriptsize\lefteqn{(\gamma)_+}}}\,\hnkl\,
\Li_2\left(e^{2\pi \I (l^\Lambda \rho_\Lambda - k_\Lambda \xi^\Lambda )} \right)\, ,
\label{prepH}
\ee
and used as a replacement for $G_{\rm A}$ in  the transition functions \eqref{symp2-inst},
which we now
treat as infinitesimal perturbations:\footnote{In this equation,
$\Hpij{0\ell_\pm}$ is a function of $\xii{0}^\Lambda$ and
$\txii{\ell_\pm}_\Lambda$, which are equal to $\xi^\Lambda$
and  $\frac{\I}{2} \rho_\Lambda$ at this order.}
\be
\label{symp3-inst}
\hHpij{0\ell_+}=-\frac{\I}{2}\,G_{\rm A/B}(\xi^\Lambda,\rho_\Lambda)\, ,
\qquad
\hHpij{0\ell_-}=-\frac{\I}{2}\,\bG_{\rm A/B}(\xi^\Lambda, \rho_\Lambda)\, .
\ee
The precise range of summation $(\gamma)_+$ in \eqref{prepH} is
left unspecified at this stage;
it must however have support on charges $\gamma=(k_\Lambda,l^\Lambda)$
with $\Re \( \Wkl \)>0$,
where %$\Wkl$ is proportional to the ``central charge'',
\be
\label{WTkl}
\Wkl \equiv \cR \left( k_\Lambda z^\Lambda - l^\Lambda F_\Lambda(z) \right)\, ,
\ee
and reproduce \eqref{rangep} when $l^\Lambda=0$.
There may be additional restrictions on the charge vector $\gamma$
generalizing the effective or anti-effective condition in  \eqref{rangep}, but we shall leave
this question open. It is also
important to note that  \eqref{prepH} was obtained by covariantizing the contributions
of instantons  with $l^\Lambda=0$,
which have a vanishing Hitchin functional; it is logically possible that the sum \eqref{phiinstfull}
may include only states related to those states by electric-magnetic duality, in particular
with a  vanishing Hitchin functional too.
At any rate, eqs.\ \eqref{prepH} and \eqref{symp2-inst} parametrize the most
general QK perturbation of the one-loop corrected metric, consistent with
integer shifts of the RR moduli $\zeta^\Lambda, \tilde{\zeta}_\Lambda$ and continuous shifts
of the NS axion $\sigma$.

Using the general results from Section \ref{subsec_lindef}, it is straightforward albeit tedious to
compute the contact twistor lines and contact potential, to first order in the perturbation \eqref{prepH}.
Using \eqref{txiqline} and \eqref{defrho}, we find that the perturbed twistor lines
in the patch $\hU_0$, most appropriate for assessing symplectic invariance,  are given by
\bse
\label{xiqlineB}
\bea
\label{xiqlineB2}
\xi^\Lambda &=& \zeta^\Lambda + \cR \left(
\varpi^{-1} z^\Lambda - \varpi \, \bz^\Lambda\right) +
\frac{1}{16\pi^2}
{\sum\limits_{\gamma}} \,\hnkl\,l^\Lambda\, \Ikl^{(1)}(\varpi)\, ,
\\
\label{txiqlineB2}
\rho_\Lambda&=&
\tzeta_\Lambda
+\cR \left( \varpi^{-1} F_\Lambda - \varpi \, \bF_\Lambda \right)
+\frac{1}{16\pi^2}{\sum\limits_{\gamma}} \,\hnkl\, k_\Lambda\, \Ikl^{(1)}(\varpi)\, ,
\\
\label{txifqlineB2}
\tilde\alpha&=& \sigma
+\cR (\varpi^{-1} W-\varpi \,\bar W) +\frac{\I\chi_X}{24\pi} \log \varpi
+\frac{\I}{2\pi^2}{\sum\limits_{\gamma}}  \, \hnkl
\(\varpi^{-1}\Wkl+\varpi\bWkl \)\Kkl
\nn\\
&& + \frac{1}{16\pi^2}
{\sum\limits_{\gamma}}  \hnkl\[\frac{1}{\pi \I}\,
\Ikl^{(2)}(\varpi)+\(\Thkl+\varpi^{-1}\Wkl-\varpi\bWkl\)\Ikl^{(1)}(\varpi) \]\, ,
\eea
\ese
where the sum over $\gamma=(k_\Lambda,l^\Lambda)$ runs over the union
of $(\gamma)_+$ and its opposite $(\gamma)_-$
(in particular, it does not include the zero class).
In \eqref{xiqlineB},  $\Wkl$  and $W$ are as  defined in  \eqref{defW} and \eqref{WTkl} with
\be
\label{xzetaB}
\begin{split}
\zeta^\Lambda \equiv A^\Lambda\, , &
\qquad
\tzeta_\Lambda \equiv B_{\Lambda}+A^\Sigma \Re F_{\Lambda\Sigma}
+\frac{1}{4\pi^2}\, \Im F_{\Lambda\Sigma}\,
{\sum\limits_{\gamma}} \,  \hnkl\,l^\Sigma
\Kkl\, ,
\\
& \sigma \equiv  - 2 B_\alpha  - A^\Lambda B_\Lambda
+\frac{1}{2\pi^2}\,A^\Lambda  \Im F_{\Lambda\Sigma}\,
{\sum\limits_{\gamma}} \,  \hnkl\, l^\Sigma
\Kkl\, ,
\end{split}
\ee
\be
\Thkl
\equiv k_\Lambda A^\Lambda - l^\Lambda\(B_\Lambda+ A^\Sigma\Re F_{\Lambda\Sigma}\)
\label{THkl}
\ee
and
\be
\begin{split}
\Kkl\equiv &
\sum\limits_{m=1}^{\infty} \frac{1}{m} \sin\(2\pi m \Thkl\)\,
K_0\(4\pi m |\Wkl|\)\, ,
\\
\Ikl^{(\nu)}(\varpi)
\equiv &
\sum_{m=1}^{\infty} \sum_{s=\pm 1} \frac{s^\nu}{m^\nu}\,
e^{-2\pi \I s m\Thkl }
\int_{0}^{\infty}\frac{\d t}{t}\, \frac{t-\epskl s\I \varpi}{t+\epskl s\I\varpi}\,
e^{-2\pi m\epskl\( t^{-1} \Wkl +t \bWkl\)} \, ,
\end{split}
\label{IKkl}
\ee
where $\epskl =\sign (\Re \Wkl)$.
Eqs. \eqref{xzetaB} generalize the relations \eqref{relABzeta} to the presence
of D-instanton corrections and ensure that
under electric-magnetic duality, $(\zeta^\Lambda, \tzeta_\Lambda)$  and
$(\xi^\Lambda,\rho_\Lambda)$ transform as a vector while $\sigma$
and $\tilde\alpha$ are invariant. Note that in the leading instanton approximation,
$\Thkl= k_\Lambda \zeta^\Lambda - l^\Lambda \tzeta_\Lambda$.
The twistor lines in other patches can be obtained by applying
the transformation rule \eqref{pert:symp}.

Finally, inserting \eqref{prepH} in \eqref{eqchip}, we obtain the perturbed contact potential,
\be
\begin{split}
e^{\Phi_{\rm A/B}} =& \frac{\cR^2}{4}\, K(z,\bz)+\frac{\chi_X}{192\pi}
\\
& +\frac{1}{8\pi^2}{\sum\limits_{\gamma}} \hnkl\sum\limits_{m> 0}
\frac{|\Wkl|}{m}\, \cos\(2\pi m \Thkl\)
K_1(4\pi m| \Wkl|)\, .
\end{split}
\label{phiinstfull}
\ee
Through \eqref{Knuflat} this result encodes the K\"ahler potential on $\cZ$.

In the absence of instanton corrections, \eqref{xiqlineB}, \eqref{xzetaB}, \eqref{phiinstfull}
reduce to \eqref{gentwi}, \eqref{relABzeta}, \eqref{phipertA},
respectively.\footnote{The apparent difference of the contact potentials by the factor of 2
is due to that the sum in \eqref{phiinstfull} goes over all lattice of charges,
including the negative ones.}
While the results
above hold in general to first order in the instanton corrections, they become exact
in the case $l^\Lambda=0$, where the toric isometries are unbroken. It would be interesting
to investigate the transformation properties of \eqref{xiqlineB} under  S-duality in this
case. S-duality should become manifest after Poisson resummation on $k_0$, but will require
correcting the tree-level action \eqref{SL2Zxi} and mirror map \eqref{physmap}.
S-duality is clearly broken by D-instanton effects, but may be recovered once NS5-brane
instantons are included. Both of these issues lie beyond the scope of this work.
Finally, note that the description of the twistor space given in this and the
preceding section is not rigorous due to the occurrence of open contours.
A more rigorous construction of the twistor space can be found in Appendix \ref{ap_rig}.

%---------------------------------------------------------
\subsection{General D-instanton corrections}
\label{secDinst}
%---------------------------------------------------------

In this section we interpret the corrections to the contact potential \eqref{phiinstfull}
as Euclidean D-brane instantons.

Using the asymptotic behavior $K_s(z)\sim \sqrt{\pi/(2z)} e^{-z}(1+\cO(1/z))$
of the modified Bessel function,
the classical instanton action associated to a general term in the sum \eqref{phiinstfull} with
$m>0$ and $(k_\Lambda, l^\Lambda)\neq 0$ is given by
\be
S_{\rm cl}=4\pi m |\Wkl|
+2\pi \I m \Thkl \, .
\label{Sin}
\ee
From the point of view of Type IIA string theory compactified on the CY
threefold $X$, instantons
should correspond to Euclidean D2-branes wrapping a special Lagrangian submanifold
in the homology class $\gamma = k_\Lambda \gamma^\Lambda - l^\Lambda \gamma_\Lambda
\in H_3(X, \IZ)$ (or more precisely, to elements in the Fukaya category $\cF(X)$, see
e.g. \cite{Aspinwall:2004jr} for a nice review). Rewriting \eqref{Sin} as
\be
S_{\rm cl}=8\pi m \sqrt{e^\phi - \frac{\chi_X}{192\pi}} \, |Z(\gamma)|
+2\pi \I m \left( k_\Lambda \zeta^\Lambda - l^\Lambda \tzeta_\Lambda \right) \, ,
\label{Sin2}
\ee
where $Z(\gamma)$ is the normalized central charge function on $H_3(X, \IZ)$,
\be
\label{defZ}
Z(\gamma) \equiv \frac{k_\Lambda z^\Lambda- l^\Lambda F_\Lambda(z)}{\sqrt{K(z,\bz)}} \, ,
\ee
and recalling that $e^{\phi/2}=1/g_4$, we recognize in the weak coupling limit $e^\phi\to\infty$
the action of $m$ Euclidean D2-branes wrapping the 3-cycle $\gamma$. The one-loop
correction proportional to $\chi_X$ can be viewed as a quantum correction to the volume
of $Y$, and was already seen for the universal hypermultiplet in
\cite{Alexandrov:2006hx}.
The infinite series of power corrections to
the exponential behavior of the modified Bessel function $K_1$
should correspond to perturbative corrections in the background of the D-instanton.
The ``instanton measure" $\hnkl$ is unknown at this stage, but presumably
counts the number of states
in $\cF(X)$ with charge $\gamma$.

On the Type IIB side,  BPS D-instantons  correspond
to elements in the derived category of coherent
sheaves $\mathcal{D}(Y)$ \cite{MR1403918,Douglas:2000gi}.
In plain (but oversimplified) terms, they
are obtained by wrapping $N$ Euclidean D$5$-branes on $Y$, and allowing a non-trivial
supersymmetric $U(N)$ gauge configuration $F$ on their worldvolume.\footnote{Instantons
with zero D5-brane charge can be obtained as bound states of D5 and
anti-D5-branes, which are also in  $\mathcal{D}(Y)$.} In the large
volume limit, their classical action is given by \cite{Minasian:1997mm,Freed:1999vc,Aspinwall:2004jr}
\be
\label{Sbbrane}
S_{\rm cl}= \tau_2 \left| \int_Y  e^{-\mathcal{J}} {\rm ch}(F) \sqrt{ {\rm td}(Y)} \right|
+ \I  \int_Y  \,A\, e^{-B} {\rm ch}(F) \sqrt{ {\rm td}(Y)}  \, ,
\ee
where ${\rm ch}$ and ${\rm td}$ denote the Chern character and Todd class,
\be
\begin{split}
{\rm ch}&=
c_0 + c_1 + \left(\frac12\, c_1^2 - c_2\right)
+\frac12\left(c_3-c_1 c_2+\frac13\, c_1^3\right)+ \dots\, ,
\\
{\rm td} &= 1 + \frac12\, c_1 + \frac{1}{12}(c_2+c_1^2) + \frac{1}{24}\, c_1 c_2 + \dots
\\
\end{split}
\ee
with $c_1(Y)=0$ by the CY condition, and $A$ is the sum of RR forms,%
\be
\label{Apot}
A = A^{(0)} + A^{(2)} +A^{(4)} + A^{(6)}  \, .
\ee
The actions \eqref{Sbbrane} and \eqref{Sin} match in the large volume limit, provided
the charges and RR scalars are identified via
\bse
\bea
{\rm ch}(F) \sqrt{ {\rm td}(Y)}  &=& l^0 + l^a\, \omega_a - k_a\, \omega^a + k_0\, \omega_Y\ ,
\label{chkl}\\
A\, e^{-B} &=& \zeta^0 - \zeta^a \, \omega_a - \tzeta_a\, \omega^a - \tzeta_0 \, \omega_Y\ .
\label{ABze}
\eea
\ese
Eq. \eqref{chkl} gives the standard relation between charges and the characteristic classes
of $F$,
\be
\label{physmap2}
\begin{split}
l^0 &= N\, ,\qquad
l^a = \int_{\gamma^a}\!\! c_1(F)\, ,
\qquad
k_a =  - \int_{\gamma_a}  \left[  \left(\frac12\, c_1^2(F) - c_2(F)\right)
+\frac{N}{24}\, c_2(Y) \right] ,
\\
k_0 &= \int_Y \left[ \frac12\left(c_3(F)-c_1(F) c_2(F)+\frac13\, c_1^3(F)\right)
+\frac{1}{24}\, c_2(Y) c_1(F) \right]
\, ,
\end{split}
\ee
while \eqref{ABze} combined with \eqref{RRiib} reproduces the mirror
map~\eqref{physmap}.

As in the Type IIA case, power corrections to the exponential behavior of $K_1$ should correspond to
perturbative corrections in the instanton background, and the ``instanton measure''
$\hnkl$ should
correspond to the number of states in $\mathcal{D}(Y)$ with
D$(-1,1,3,5)$ charges  $(k_0, k_a,l^a,l^0)$ in $H_{\rm even}(Y)$,  possibly with a restriction
on the allowed charges.

%%%%%%%%%%%%%%%%%%%%%%%%%%%%%%%%%%%%%%%%%%%%%%%%%%%%%%%%%%%%%%%%%%%%%%
\subsection{Exact twistor space in presence of D-instantons \label{secrig}}
%%%%%%%%%%%%%%%%%%%%%%%%%%%%%%%%%%%%%%%%%%%%%%%%%%%%%%%%%%%%%%%%%%%%%%

We now suggest a construction of the twistor space $\cZ$
in presence of D-instanton corrections,
essentially identical to the one given in  \cite{Gaiotto:2008cd} in the gauge theory
context, which should be exact
in the absence of NS5-brane instantons.

As in \cite{Gaiotto:2008cd}, each charge vector
$\gamma=(k_\Lambda, l^\Lambda)$ defines a pair of ``BPS rays''
$\ell_\pm (\gamma)$ on $\CP$ and two hemispheres $V_\pm(\gamma)$ defined by
\be
\ell_\pm (\gamma)= \{ \varpi :\,  \pm \Wkl/\varpi \in \I\IR^{-} \}\, ,
\qquad
V_\pm (\gamma)= \{ \varpi : \, \pm\Im(\Wkl/\varpi) <0 \}\, ,
\label{lpmGMN2}
\ee
in such a way that $|e^{\mp \I (k_\Lambda \xi^\Lambda-l^\Lambda\rho_\Lambda)}|<1$
in $V_\pm(\gamma)$,
and that $e^{\mp \I (k_\Lambda \xi^\Lambda-l^\Lambda\rho_\Lambda)}$ is exponentially
suppressed at $\varpi\to 0$ and $\varpi\to \infty$ in $V_\pm$.
We propose that across all BPS rays
$\ell_\pm(\gamma)$ the complex contact structure experiences
finite contact transformations $U_\gamma$ generated by
\be
\label{elcon}
\hSij{ij}_\gamma( \xii{i}^\Lambda,\txii{j}_\Lambda,\ai{j})
 = \ai{j}+ \xii{i}^\Lambda \,  \txii{j}_\Lambda + \frac{\I}{2(2\pi)^2}\, \hnkl\,
\Li_2\left(e^{\mp 2\pi \I (k_\Lambda \xii{i}^\Lambda+2\I l^\Lambda \txii{j}_\Lambda   )} \right)\, .
\ee
Actually, the precise angular location of the BPS rays $\ell_\pm(\gamma)$
where the contact transformation is performed
is unimportant, provided they stay inside the hemispheres $V_\pm(\gamma)$, respectively, and the
angular order between BPS rays of different charges is preserved.  Thus, one may ``pile up'' the
BPS rays up in just two composite rays located on the positive and negative
imaginary axis \cite{Gaiotto:2008cd}. Across these two rays, the contact structure experiences
the product of all elementary contact transformations \eqref{elcon}, ordered
counterclockwise according to the phase of the central charge $\Wkl$:
\be
\label{Spm}
U_+ = \prod^\ccwarrow_{\Re(\Wkl)>0} U_\gamma\, ,
\qquad
U_- = \prod^\ccwarrow_{\Re(\Wkl)<0} U_\gamma\, ,
\ee
where the product denotes the composition of contact transformations.
The latter can be computed from the generating functions  using \eqref{compS}.
Together with the contact transformations \eqref{symp-cmap} determining the
perturbative part of the hypermultiplet metric, this defines a twistor space $\cZ$
which should provide the exact metric on the hypermultiplet moduli space $\cM$
in the absence of NS5-brane instantons.

The ordering of the BPS rays depends on the moduli $z^\Lambda$ via the central
charge function \eqref{defZ}, and changes across lines of marginal stability (LMS)
where the phase of the central charges of two BPS instantons $\gamma_1$ and
$\gamma_2$ become aligned. At the same time, the value of the invariants
$\hnkl$ is expected to change, in such a way that the
products $S_+$ and $S_-$ stay invariant. As explained in    \cite{Gaiotto:2008cd}
and further discussed in Section \ref{secwall}, this consistency condition is identical in form
to the wall-crossing formula for generalized Donaldson-Thomas invariants found
in \cite{ks}. Thus, it strongly suggests that the instanton measure
should be identified to these generalized Donaldson-Thomas invariants.

In the leading instanton approximation, the infinite products  in \eqref{Spm}
reduce to an infinite sum, and the contact transformations $S_\pm$ are generated by the
functions $G_{\rm A/B}$ and $\bar G_{\rm A/B}$ as described in Section \ref{linins}.

%%%%%%%%%%%%%%%%%%%%%%%%%%%%%%%%%%%%%%%%%%%%%%%%%%%%%%%%%%%%%%%%%%%%%%%%%%%%
\section{Discussion}
%%%%%%%%%%%%%%%%%%%%%%%%%%%%%%%%%%%%%%%%%%%%%%%%%%%%%%%%%%%%%%%%%%%%%%%%%%%%
\label{sec_disc}

In this work, we have studied D-instanton corrections to the hypermultiplet branch $\cM$ of Type II
compactifications on a Calabi-Yau threefold using twistor techniques. Our main result is the
instanton-corrected ``contact potential''  \eqref{phiinstfull}, which, together with the ``contact
twistor lines''  \eqref{xiqlineB} provides sufficient information to determine the instanton-corrected
QK metric on the hypermultiplet moduli space, in the ``leading instanton" approximation. These
results follow from a simple deformation of the complex contact geometry on the twistor space $\cZ$
of $\cM$, controlled by the holomorphic function \eqref{prepH} (more accurately, a section of
$H^1(\cZ,\cO(2))$. In Section \ref{secrig}, we have proposed how this perturbation could
be elevated to a finite deformation of the twistor space $\cZ$, which should yield the exact
QK metric on $\cZ$ in the sector without NS5 branes.
In the remainder of this work, we comment on some possible relations of these results to
the counting of 4D BPS black holes,  and to the wall-crossing formula
of Kontsevich and Soibelman, and speculate on the
form of the NS5-brane instanton corrections.

%---------------------------------------------------------
\subsection{Instanton corrections and black hole partition functions \label{bhsec}}
%---------------------------------------------------------

In \cite{Gunaydin:2005mx}, it was suggested on general ground that instanton-corrected
BPS couplings in three dimensions may provide a useful packaging for the BPS black hole
degeneracies in four dimensions. Here, we apply these general ideas to the case of
$\cN=2$ supersymmetry, and argue that the instanton measure $\hnkl$
in Type IIB (resp. Type IIA) string theory compactified on $Y$ is directly
related to the microscopic indexed
degeneracy of 4D black holes in Type IIA (resp. Type IIB) string theory compactified on the
{\it same} Calabi-Yau threefold $Y$.

For this purpose, consider the compactification of Type IIB string theory down
to  three dimensions on the product of $Y$ times
a circle of radius $R=e^{U} l_P$, where $l_P$ is the 4D Planck length.
The moduli space in three dimensions
factorizes into the product
\be
\label{prodmod}
\cM_3= \cM_{\rm HM}^B \times \cM_{\rm VM}^B
\ee
of two QK manifolds,
of dimension  $4(h^{1,1}(Y)+1)$ and $4 (h^{1,2}(Y)+1)$, respectively.

The first factor $\cM_{\rm HM}^B$ is independent of the radius $R$ (since
vector multiplets and neutral hypermultiplets are decoupled
at two-derivative order), and coincides with the hypermultiplet moduli space $\cM_{\rm HM}^B$
in four dimensions. The latter was described at the perturbative level in Section \ref{seciib}, and
receives instanton corrections from Euclidean D-branes wrapping supersymmetric cycles
in $H_{\rm even}(Y)$ as found in Section \ref{secDinst}.

On the other hand, the second
factor $\cM_{\rm VM}^B$  contains the
radius $R/l_P$, the complex structure of $Y$,
the electric and magnetic Wilson lines $(\tzeta_\Lambda,\zeta^\Lambda)$
of the graviphoton and vector multiplets in $D=4$, and the NUT scalar $\sigma$
(dual to the off-diagonal part
of the metric). In the limit $R\gg l_P$, it is given by the $c$-map of the complex
structure moduli space
$\cM_{\rm cs}(Y)$.
At finite $R$, $\cM_{\rm VM}^B$ is expected to receive loop corrections from Kaluza-Klein
states running around the Euclidean circle, and instanton corrections
from 4D BPS black holes of charge $\gamma=(p^\Lambda, q_\Lambda)$
whose worldline winds around the
circle.\footnote{Analogous contributions of 4D monopoles
to the 3D effective potential are famously responsible
for the confinement of $D=2+1$ compact Maxwell theory  \cite{Polyakov:1976fu}. The reason
that only BPS black holes can contribute to the metric is the standard saturation of fermionic
zero-modes, see e.g. \cite{Kiritsis:1999ss}.}
The classical action of these configurations is given by the mass of 4D black hole
times the length of the circle, plus the coupling to the Wilson lines,
\be
\label{Sbh}
S_{\rm cl} = 2\pi e^U | Z(\gamma)  |+2\pi \I (\zeta^\Lambda q_\Lambda
- \tzeta_\Lambda p^\Lambda )\, ,
\ee
where the central charge
$Z(\gamma)$ (a function of the vector multiplet moduli and the black hole charges)
takes the same form as in  \eqref{defZ}.
In addition to these $\sigma$-independent contributions, there are also Euclidean
configurations with NUT charge $k\neq 0$, inducing terms proportional
to $e^{\I k  \sigma}$ in the low-energy effective action.
Similarly, in Type IIA compactified on $Y\times S^1$ the moduli space
takes the product form $\cM_3= \cM_{\rm VM}^A \times \cM_{\rm HM}^A$,
with the role of \kahler and complex structure moduli
being exchanged.

It is well-known that T-duality along the circle exchanges
the two factors in the three-dimensional moduli space \eqref{prodmod} \cite{Seiberg:1996ns},
\be
\cM_{\rm HM}^B = \cM_{\rm VM}^A \, ,
\qquad
\cM_{\rm VM}^B = \cM_{\rm HM}^A\, ,
\ee
in particular it exchanges the radius $U$ with the four-dimensional dilaton $\phi$.
This implies that (i) the one-loop correction on the hypermultiplet branch, proportional to
$\chi_Y$, should reproduce the loop corrections from KK states on the vector multiplet branch,
and (ii) the D-instanton contributions to $\cM_{\rm HM}^B$ (already present in $D=4$) should be
mapped to black hole instantons contributions to  $\cM_{\rm VM}^A$ (arising in the
compactification to $D=3$). While we have not attempted to check (i), it is clear that the
classical actions \eqref{Sin2} and \eqref{Sbh} agree  provided T-duality exchanges
\be
e^U \leftrightarrow 4 \sqrt{e^\phi - \frac{\chi_X}{192\pi}}\, ,
\ee
implying a one-loop correction to the usual $c$-map.

At this point, it may be worthwhile to note that the correction
terms in the contact potential \eqref{phiinstfull}
are identical in form to the radial wave function for BPS black holes computed in
\cite{Neitzke:2007ke,Gunaydin:2007bg},
except for the one-loop correction proportional to $\chi_X$. This is hardly surprising,
since in the context of $D=4, \cN=2$ supergravity, spherically symmetric
BPS instanton configurations are described by the same geodesic
motion which controls the radial profile of BPS
black holes \cite{Gutperle:2000ve,Gunaydin:2005mx}.
The radial quantization of BPS black hole solutions leads to a
quantum Hilbert space of functions which happens to coincide with the space
$H^1(\cZ,\cO(2))$ of QK deformations of $\cM$.  This has an important
practical consequence: for a fixed value of the moduli $z^\Lambda$, the
instanton configurations which dominate the sum  \eqref{phiinstfull} are given by
extremizing the central charge $|Z(\gamma)|$ with respect to the charges $\gamma$.
It would be interesting to investigate this ``reverse attractor mechanism"  further.

While the action of D-instantons and black holes are easily matched, the relation between
the summation measures is more subtle. In discussing this issue, it is useful to bear in mind
an analogous but simpler problem, namely the relation between the D$(-1)$-instanton measure for $R^4$
couplings in $D=10$ Type IIB string theory \cite{Green:1997tv}, and the Witten index of
D0-branes in $D=10$ Type IIA string theory \cite{Green:1997tn}. Since $N$ D0-branes
have a single bound
state at threshold for any $N>0$, the index should be equal to 1. The Witten index
is given by a functional integral in $U(N)$ supersymmetric quantum mechanics with
16 supercharges, with periodic boundary conditions for the fermions along the
Euclidean time circle of length $\beta$. However, due to flat directions in the potential,
only the low temperature limit $\beta\to\infty$ is expected to yield the Witten index $\Omega(N)$.
On the other hand, the D$(-1)$-instanton measure $\mu(N)$ is given by a $U(N)$ matrix integral, i.e.
the reduction of the quantum mechanics on a circle of vanishing size $\beta\to 0$.
After regulating volume divergences, the difference
\be
\Omega(N) - \mu(N) = \int_0^{\infty} \de\beta\, \frac{\pa}{\pa\beta}
\Tr\left[ (-1)^F e^{-\beta H} \right]\, ,
\ee
rewritten as a ``bulk" contribution to the index \cite{Yi:1997eg,Sethi:1997pa},
was evaluated in  \cite{Green:1997tn} and found to agree with the answer predicted
by S-duality \cite{Green:1997tv}, $\mu(N)=\sum_{d|N} 1/d^2,\ \Omega(N)=1$. In particular, when $N$ is
a prime number, the instanton measure and the Witten index agree.

This analogy suggests  that in the absence of marginal directions
in the potential, i.e. for non-threshhold bound states, the instanton measure
$\hnkl$ for Type IIB/$Y$ and the indexed degeneracy
$\Omega(k_\Lambda,l^\Lambda)$ of 4D BPS black holes in Type IIA/$Y$
should agree (with a similar statement upon exchanging Type IIA and IIB). Thus,
the metric on the hypermultiplet branch appears to be a very convenient packaging
for the indexed degeneracies of 4D BPS black hole in the dual theory. In particular,
it gives a natural way to encode the dependence of the black hole spectrum
on the values of the moduli at spatial infinity, as we now discuss.

%---------------------------------------------------------
\subsection{Wall crossing  \label{secwall}}
%---------------------------------------------------------

The spectrum of single-particle states is known to jump across lines
of marginal stability (LMS), where the phase of the central charge
of two BPS states align. This phenomenon has been much studied
in the context of $\cN=2$ supersymmetric gauge theories
(see e.g. \cite{Argyres:1995gd,Ferrari:1996sv,Bilal:1996sk}), and also
takes place in $\cN=2$ supergravity theories, where it has a macroscopic
description in terms of multi-centered black hole configurations \cite{Denef:2000nb}:
as the LMS is approached from one side, the distance between the centers
diverges and the configuration becomes unbound.
On the other side of the LMS, the bound state no longer exists
as a single-particle state, but it is replaced by a continuum of multi-particle states with
the same total charge.
Thus, by analyzing the leading instanton contributions at a given point on the 3D moduli space
in the large radius limit $U\to\infty$, one should be able to determine the BPS spectrum
at that particular point. Moreover, since no massless state typically occurs on the LMS,
the hypermultiplet  metric is expected to be smooth, with the single instanton contribution
on one side of the LMS matching the multi-instanton contribution on the other side.
This should provide strong constraints on the discontinuity of the one-particle BPS
spectrum across the LMS.

This idea was demonstrated recently in the context of
rigidly supersymmetric gauge theories with 8 supercharges in 4
dimensions \cite{Gaiotto:2008cd}. In particular, the authors showed that the
\hk moduli space $\cS$ of the gauge theory compactified down to three dimensions
gives a natural physical setting for the
Kontsevich-Soibelman wall-crossing formula \cite{ks,moorelec}:  the latter ensures
that the leading instanton effects on the twistor space  combine with each other consistently
so as to produce a regular HK manifold. To see why this may be true, recall that
on very general ground, the  ``generalized Donaldson-Thomas invariants'' $\Omega(\gamma)$
must satisfy  \cite{ks,moorelec,Gaiotto:2008cd}
\be
\label{wc}
\prod^\ccwarrow_{\substack{\gamma=n \gamma_1+m \gamma_2\\m>0, n>0} }
U_{\gamma}^{\Omega^-(\gamma)} =
\prod^\cwarrow_{\substack{\gamma=n \gamma_1+m \gamma_2\\m>0, n>0} }
U_{\gamma}^{\Omega^+(\gamma)}\, ,
\ee
where $\Omega^{-}(\gamma)$ and $\Omega^{+}(\gamma)$ denote the value of $\Omega(\gamma)$ on
either side of the LMS where the phases of $Z(\gamma_1)$ and $Z(\gamma_2)$, align. Here
\be
\label{wcu}
U_{\gamma} \equiv \exp\left( \sum_{n=1}^{\infty} \frac{1}{n^2}\, e_{n\gamma} \right)\ ,
\ee
where $e_\gamma\equiv e_{p,q}$ are generators of  the Lie algebra
\be
\label{algks}
\left[ e_{p,q}, e_{p',q'} \right] = (-1)^{ p^\Lambda q'_\Lambda - p'^\Lambda q_\Lambda}
\,\left(p^\Lambda q'_\Lambda - p'^\Lambda q_\Lambda \right)\, e_{p+p',q+q'}\, .
\ee
Except for the sign $(-1)^{ p^\Lambda q'_\Lambda - p'^\Lambda q_\Lambda}$,
which can be absorbed into a redefinition of  $e_{p,q}$ by a choice of
``quadratic refinement" \cite{Gaiotto:2008cd}, this is the algebra of infinitesimal
symplectomorphisms on the complex torus $ (\IC^\times)^{2n}$,
where $e_{p,q} = e^{2\pi \I(q_\Lambda \xi^\Lambda - p^\Lambda \txi_\Lambda)}$
is a basis of contact Hamiltonians and the commutator
is the Poisson bracket $\[\mu_1,\mu_2\]=(\I/2\pi)(
\pa_{\xi^\Lambda}  \muh_1 \pa_{\txi_\Lambda}  \muh_2-
\pa_{\xi^\Lambda}  \muh_2 \pa_{\txi_\Lambda}  \muh_1)$. Indeed,
this complexified torus  can be identified as the twistor space $\cZ_\cS$
of the HK manifold $\cS$, and the relation \eqref{algks} guarantees the
consistency of the symplectic structure across the LMS \cite{Gaiotto:2008cd}.

Returning to the case of Type IIB string theory compactified on $Y$,
where the moduli space $\cM$ is QK rather than HK, it is natural to expect that a similar construction
operates at the level of the twistor space $\cZ$ equipped with its complex contact
structure.  Indeed, by using the requirement of  S-duality invariance, we have found that
in the ``leading instanton" approximation,  instanton corrections induce
contact transformations generated by the sum of dilogarithms \eqref{prepH}.
The latter originated by Poisson resummation from the trilogarithms present
in the worldsheet instanton sum \eqref{lve}. The
occurrence of the same   dilogarithm function in \eqref{prepH} as in \eqref{wcu}
gives a strong hint that the instanton measure $\hnkl$
should be identified with the generalized Donaldson-Thomas invariants of \cite{ks},
and in turn with the index degeneracies of 4D BPS black holes.

Alas, this sequence of identifications raises serious puzzles:
the indexed degeneracies $\Omega(\gamma)$ of large black holes
are known to grow  exponentially as  $\Omega\sim e^{\lambda^2}$ when the charge
vector $\gamma$ is rescaled
by a common factor $\lambda$, while the exponential of the classical action decreasing
only as $e^{-\lambda}$. Assuming that the instanton measure
were equal to the black hole degeneracy, it would seem impossible that the instanton
sum could converge at all.\footnote{For similar reasons, the 6-derivative BPS couplings
in 3D string vacua with 16 supercharges, or for 14-derivative couplings in
3D vacua with 32 supercharges, may be ill defined.} It is conceivable however that
the instanton measure, and indeed the generalized Donaldson-Thomas invariants,
may have support on ``polar''
states \cite{Denef:2007vg} (i.e. states with imaginary entropy in the
supergravity approximation), whose degeneracies
grow less rapidly. Another puzzle is the absence of
quantum corrections to the hypermultiplet  moduli space metric in
Type II compactifications on certain self-mirror CY manifolds~\cite{Ferrara:1995yx},
which are nevertheless expected
to have a non-trivial spectrum of BPS black holes.
It would be interesting to check whether
black holes or instantons in these models have accidental
fermionic zero-modes which forbid their contribution to the
index and/or to the metric.

%---------------------------------------------------------
\subsection{NS5-brane instantons}
%---------------------------------------------------------

We now briefly comment on the effects of NS5-branes on the hypermultiplet metric.\footnote{These
are dual to the effects of Euclidean configurations with non-zero NUT charge on the vector
multiplet branch in 3 dimensions mentioned below Eq. \eqref{Sbh}.}
Since these instanton
configurations carry magnetic charge under the NS two-form $B$, they must break
the shift isometry along the NS axion direction to a discrete subgroup. Since
the NS axion enters linearly in the complex coordinate
$\tilde\alpha=4\I \ai{0} +2 \I \txii{0}_\Lambda \xii{0}^\Lambda=\sigma+\dots$, such corrections
must take the form, at the infinitesimal level,
\be
\hHpij{ij}(\xii{i}^\Lambda, \txii{j}_\Lambda,\ai{j}) \sim
\exp\left(-4k \, \ai{j} \right) H^{[ij]}_{{\scriptscriptstyle\smash{(1)}},k}
(\xii{i}^\Lambda, \txii{j}_\Lambda)\, ,
\ee
where $k$ is the NS5-brane charge. This causes some technical difficulty,
since $\hHpij{ij}$ is no longer  independent of $\ai{j}$, and the QK geometry can
no longer be described by a single $\varpi$-independent contact potential.

A more conceptual problem however is the fact that for
non-vanishing NS5-brane charge $k$,
the translations along the RR axionic directions no longer commute. Instead,
they generate a Heisenberg algebra
\be
\[ P^\Lambda, Q_\Sigma \] = -2 \delta^\Lambda_\Sigma K\, ,
\ee
where
\be
P^\Lambda=\pa_{\tzeta_\Lambda}-\zeta^\Lambda\pa_\sigma\, ,
\qquad
Q_\Lambda=-\pa_{\zeta^\Lambda}-\tzeta_\Lambda\pa_\sigma\, ,
\qquad
K=\pa_\sigma\, ,
\ee
with the effective Planck constant $K$ being proportional to the NS5-brane charge $k$.
Thus, a Fourier decomposition such as \eqref{prepH} is no longer applicable. Instead,
the plane wave solutions
appearing in \eqref{prepH} should be replaced by wave functions of a
charged particle on a torus with magnetic flux $k \de\xi^\Lambda\wedge \de\rho_\Lambda$.
It is tempting to speculate that the coefficients of this non-Abelian Fourier
decomposition may be related to  the  ``quantum invariants" defined in \cite{ks},
with the classical dilogarithm $\Li_2$ being replaced by its quantum version.

In the absence of an obvious guess for the form of these NS5-corrections, one may
consider the longer route proposed in  \cite{RoblesLlana:2007ae}:
by mirror symmetry, the B-type D2-brane instantons in Type IIA are mapped to D3- and
D5-brane instantons. A further use of S-duality in principle would map D5-brane
instantons to NS5-brane instantons in Type IIB, and finally to NS5-brane instantons in
Type IIA  via mirror symmetry. Given the complexity of the transformation rules \eqref{SL2Zxi}
of the twistor lines at tree-level, it is a challenging problem to covariantize the B-type instanton
contributions under S-duality. We hope to return to this issue in a forthcoming publication.

\acknowledgments
We are grateful to M. Haack, N. Halmagyi, M. Kontsevich, D. Morrison,
G. Moore, A. Neitzke,  and S. Stieberger for valuable discussions.
The research of S.A. is supported by CNRS and by the contract
ANR-05-BLAN-0029-01. The research of B.P. is supported in part by ANR(CNRS-USAR)
contract no.05-BLAN-0079-01. F.S.\ acknowledges financial support from the ANR grant
BLAN06-3-137168. S.V. thanks the Federation de Recherches ``Interactions Fondamentales''
and LPTHE at Jussieu for hospitality and financial support.
Part of this work is also supported by the EU-RTN network MRTN-CT-2004-005104
``Constituents, Fundamental Forces and Symmetries of the Universe''.

\appendix

%%%%%%%%%%%%%%%%%%%%%%%%%%%%%%%%%%%%%%%%%%%%%%%%%%%%%%%%%%%%%%%%%%%%%%
\section{More twistor constructions}
%%%%%%%%%%%%%%%%%%%%%%%%%%%%%%%%%%%%%%%%%%%%%%%%%%%%%%%%%%%%%%%%%%%%%%
\label{ap_GMN}

In this appendix, we revisit the twistor space formulation
of the moduli space $\cS$ underlying $\cN=2$ gauge theories on  $\IR^3\times S^1$  in \cite{Gaiotto:2008cd},
focusing on the
case of $SU(2)$ gauge group without matter for simplicity. The  hyperk\"ahler metric on $\cS$
resulting from integrating out  one BPS particle of electric charge $q>0$  winding around the circle
was constructed in \cite{Ooguri:1996me,Seiberg:1996ns}. In Section \ref{secap1},
we construct a set of local coordinates on its twistor space $\cZ_\cS$ which cover the whole $\CP$,
including the north and south pole where the coordinates introduced in \cite{Gaiotto:2008cd}
have an essential singularity. This is important since in the absence of such a covering, it
is not clear (to us) why the holomorphic symplectic form $\Omega(\zeta)$ should be a global
$\cO(2)$ section.

As it turns out, the symplectomorphisms underlying this construction
are essentially identical to the contact transformations
which determine the instanton-corrected twistor space
for the hypermultiplet branch discussed in Section \ref{sec:Dinst}. In Section \ref{ap_rig},
we use this observation to provide a rigorous construction of the twistor space
of the hypermultiplet moduli space in Type II compactifications
in the leading instanton approximation.

\subsection{The Ooguri-Vafa metric revisited\label{secap1}}

The authors of \cite{Gaiotto:2008cd} parametrize the real twistor lines on the twistor space
$\cZ_\cS$ of $\cS$ by
\be
\begin{split}
\xi(\zeta)& = \theta_e - \I \pi R (a/\zeta + \zeta \bar a) \, ,
\\
\txi(\zeta) & = \theta_m- \I \pi R (F_a/\zeta + \zeta \bar F_{\bar a}) + \delta\txi\, ,
\label{txiGMN}
\end{split}
\ee
where $\xi,\txi,\zeta$ are complex coordinates on $\cZ_\cS$ such that the complex
symplectic form takes the Darboux form $\Omega=\de\xi \wedge \de\txi$ (up to
an overall normalization), and $a,\bar a,\theta_e,\theta_m$ are coordinates on $\cS$.
Here,
\be
F_a \equiv \pa_a F\, ,
\qquad
F(a)\equiv \frac{q^2}{4\pi \I} \left( a^2 \log \frac{a}{\Lambda}
- \frac32\, a^2 \right)\, ,
\ee
where $\Lambda$ is the QCD scale, and
\be
\label{deltaxi}
\delta\txi = \frac{q}{2\pi} ( \CI_+ - \CI_-)\, ,
\qquad
\CI_\pm = \frac12\int_{\ell_\pm} \frac{\de\zeta'}{\zeta'} \frac{\zeta'+\zeta}{\zeta'-\zeta}
\log\left( 1- e^{\pm\I q \xi(\zeta')} \right) \, ,
\ee
whereas $\ell_\pm$ are semi-infinite lines from 0 to $\infty$, lying in the middle of the half planes $V_\pm$,
\be
\ell_\pm = \{ \zeta :\, \pm a/\zeta \in \IR^{-} \}\, ,
\qquad
V_\pm = \{ \zeta : \,\pm \Re(a/\zeta) <0 \}\, .
\label{lpmGMN}
\ee
These conditions guarantee that $|e^{\pm \I q \xi(\zeta)}|<1$ in $V_\pm$, and
that $e^{\pm \I q \xi(\zeta)}$ is exponentially suppressed at $\zeta\to 0$ and
$\zeta\to \infty$ in $V_\pm$. On the axis separating $V_+$ and $V_-$,
$e^{\I q \xi(\zeta)}$ has modulus one, and equals one for
infinitely many values of $\zeta$ accumulating near $\zeta=0$ and $\infty$.

\subsubsection*{Analytic properties of complex coordinates}

Our aim is to provide a regular set of coordinates on the twistor space defined above.
The coordinates \eqref{txiGMN} do not cover
the whole $\CP$, since the expression \eqref{deltaxi} is analytic only away from the contours of
integration $\ell_\pm$. Instead,  \eqref{txiGMN} defines two different functions $\delta\txi_>$
and $\delta\txi_<$ analytic on the half plane
$V_>$ and $V_<$, respectively, with
\be
V_{>} =  \{ \zeta :\, \Im(a/\zeta) >0 \}\, ,
\qquad
V_{<} =  \{ \zeta :\, \Im(a/\zeta) <0 \}\, .
\ee
Moreover,  under the map $\zeta\mapsto a/(\bar{a} \zeta)$,
which exchanges these two regions,
\be
\label{antipodi}
\CI_\pm( \zeta )  + \CI_\pm\left( \frac{a}{\bar a\,\zeta}\right)= 0\, ,
\qquad
\delta\txi_> ( \zeta )  + \delta\txi_< \left( \frac{a}{\bar a\,\zeta}\right) = 0\, .
\ee
Using the invariance of $e^{\I q \xi(\zeta)}$ under this map,
we may also rewrite \eqref{deltaxi} as an integral over the variable $\xi$,
\be
\begin{split}
\CI_\pm
&= \int_{\pm\I\infty}^{\xi(\mp\sqrt{a/\bar a})}\frac{\de \xi'}{\xi'-\xi}\,
\[\frac{\zeta\p_\zeta\xi}{\zeta'\p_{\zeta'}\xi'}\]
\log\left( 1- e^{\pm\I q \xi} \right)\, ,
\end{split}
\label{reprIpm}
\ee
where the factor in the square brackets is understood as a function of $\xi$ and $\xi'$.
Starting from \eqref{deltaxi}, a direct computation establishes the partial differential equations
\bse
\bea\label{A.8}
\left( \pa_a \pa_{\bar a} + \pi^2 R^2  \pa_{\theta_e}^2 \right) \CI_\pm &=& 0\, ,
\\
\left( a \pa_a - \bar a \pa_{\bar a} + \ze \pa_\ze \right)  \CI_\pm &=& 0\, .
\eea
\ese
Note that any function of $\xi$ is a solution of these equations. Moreover, any solution
of the system can be written as $F(a,\bar a,\theta_e,\zeta) = \Phi(a/\zeta,\bar a \zeta,\theta_e)$ where
$\Phi(a,\bar a,\theta_e)$ is a solution of \eqref{A.8}. The latter has a basis of
solutions $K_0(2\pi R q m |a|)\, e^{\I m q \theta_e}$ with $m\neq 0$.

The function $\delta\txi_>$  can be analytically continued into $V_+\cap V_< $
across the contour $\ell_+$.
Similarly, $\delta\txi_<$ can be analytically continued into $V_- \cap V_>$
across the contour $\ell_-$.
On their common domain of definition, the two functions
differ by the residue at $\zeta'=\zeta$ in \eqref{deltaxi},
\bse
\bea
\delta\txi_> - \delta\txi_< &=& \I q \log\left( 1- e^{\I q \xi} \right)\, ,
\qquad
\zeta \in V_+ \, ,
\label{disc1}
\\
\delta\txi_> - \delta\txi_< &=&  \I q \log\left( 1- e^{-\I q \xi} \right)\, ,
\qquad
\zeta \in V_-Ê\, .
\label{disc2}
\eea
\ese
In particular, starting from the lower left quadrant $V_+ \cap V_>$, one may analytically
continue $\delta\txi_>$ across $\ell_+$ into the upper left quadrant $V_+ \cap V_<$, picking
\eqref{disc1}, then
into the upper right quadrant $V_- \cap V_<$, picking an additive constant $2\pi q m_<, \ m_<\in \IZ$,
then into the lower right quadrant $V_- \cap V_>$ across $\ell_-$, picking \eqref{disc2}, and
back again to the  lower left quadrant, picking an extra additive constant $2\pi q m_>, \ m_>\in \IZ$.
Using the fact that
\be
\I q \log\left( 1- e^{-\I q \xi} \right) -  \I q \log\left( 1- e^{\I q \xi} \right)
= q^2 \xi\quad \mod 2\pi q\, ,
\ee
we conclude that the monodromy of $\delta\txi_>$ around $\zeta=0$ is given by
\be
\delta\txi_> (\zeta e^{-2\pi \I}) = \delta\txi_>(\zeta) + q^2 \xi \quad \mod 2\pi q\, .
\ee
Despite the fact that the part of the monodromy linear in $\xi$ can be canceled by adding
$\frac{q^2}{2\pi \I}\, \xi \log \xi$ to $\delta\txi_>$, this does not give a regular function
near $\zeta=0$ because the resulting combination still has an essential singularity at this point.

To understand how regular coordinates can be defined, let us study the
behavior of \eqref{deltaxi} near $\zeta=0$.
Taylor expanding the rational function, Fourier expanding the logarithm and
setting $\zeta'=\mp t/\bar a$ leads  to
\be
\CI_\pm = - \int_0^\infty \frac{\de t}{t} \left ( \frac12+
\sum_{k=1}^{\infty} \left( \mp \frac{\bar a \zeta}{t}
\right)^k \right) \left( \sum_{m=1}^{\infty} \frac{1}{m}\,
e^{\pm\I m q \theta_e - m \pi q R (t + |a|^2/t) } \right)\, .
\label{intIpm}
\ee
Exchanging the two sums with the integral over $t$ gives
\be
\CI_\pm = -   \sum_{m=1}^{\infty} \frac{1}{m}\, e^{\pm \I m q \theta_e}
\left[ K_0(2\pi q R m |a|) + 2 \sum_{k=1}^{\infty}
\left( \mp \zeta \sqrt{\frac{\bar a}{a}}\right)^k  K_k (2\pi q R m |a|) \right] \, .
\ee
This Taylor series correctly reproduces the limit of $\CI_\pm$ and of all its derivatives at $\zeta\to 0$
(irrespective of the direction of approach), in particular
\be
\CI_\pm(0) = -\sum_{m=1}^{\infty} \frac{1}{m}\, e^{\pm\I m q \theta_e}  K_0(2\pi q R m |a|) \, .
\ee
However the radius of convergence in the $\zeta$
variable is zero. This reflects the existence of an essential singularity at $\zeta=0$, and
the fact that the analytical continuation of $\CI_+$ across $\ell_+$
diverges when $\zeta=0$ is approached from $V_-$.
In order to expose the behavior at $\zeta=0$, one may omit the term linear in $t$ in the
exponent of \eqref{intIpm} (after subtracting the $\zeta$-independent term).
The integral over $t$ is now of Gamma function type, leading to
\be
\CI_\pm - \CI_\pm(0) \sim - \sum_{m=1}^{\infty}  \frac{1}{m}\, e^{\pm\I m q \theta_e}
\sum_{k=1}^{\infty}  \Gamma(k)
\left( \mp\frac{\zeta}{m \pi q R a } \right)^k\, .
\ee
The (divergent) sum over $k$ is recognized as the asymptotic expansion
$\e^{-z} \Ei(z)  = \sum_{k=1}^\infty (k-1)! z^{-k}$ 
valid away from the positive $z$ axis, of the Exponential Integral
$\Ei(z)\equiv \dashint_{-\infty}^z e^t \de t/t$ at $z\to \infty$.
Therefore, we conclude that the analytic behavior of $\CI_\pm$ near the origin
is characterized as
\be
\CI_\pm - \CI_\pm(0) \sim - \sum_{m=1}^{\infty}  \frac{1}{m}\, e^{\pm\I m q \theta_e}
e^{ \pm m \pi q R a / \zeta} \, \Ei\left( \mp\frac{m \pi q R a }{\zeta} \right) \, .
\ee

An important fact is that this behavior depends on $\zeta,a,\theta_e$
through the complex coordinate $\xi$ only. Indeed, using
$\xi\sim \theta_e-\I \pi R a/\zeta$ at $\zeta=0$, one finds
\be
\CI_\pm - \CI_\pm(0) \sim - \sum_{m=1}^{\infty}  \frac{1}{m}\, e^{\pm \I m q \xi}
\Ei\left( \mp\I m q \xi \right)
= \int_0^{\pm\I\infty} \frac{\de\xi'}{\xi-\xi'} \,\log\(1-e^{\pm \I q \xi'}\)  \, ,
\label{Ipmana0}
\ee
where we used the integral representation
\be
e^{-x y } \Ei(x y ) = \int_0^\infty \frac{e^{-x t}}{y- t}\, \de t
\ee
valid for $x>0$ with identifications $x=m q,\ y=\mp \I \xi,\ t = \mp \I \xi'$,
and performed the sum over $m$.
This makes it manifest that \eqref{Ipmana0} and \eqref{reprIpm} have the same
discontinuity across the integration contours.

In total, these results imply the following asymptotic behavior of $\txi$ at $\zeta=0$,
\be
\label{asdxi}
\txi(\zeta) \mathop{\sim}\limits_{\zeta\to 0}
\frac{q^2}{2\pi \I}\(\xi\log\frac{\I\zeta\xi}{\pi R\Lambda}-\xi\)
+ \frac{q}{2\pi} \sum_{m\ne 0} \frac{1}{m} \,e^{-\I m q \xi}
\Ei\left( \I m q \xi \right)\, .
\ee
The singular behavior at $\zeta=\infty$ may be studied in the same way,
or inferred from \eqref{antipodi}:
\be
\txi(\zeta) \mathop{\sim}\limits_{\zeta\to \infty}
-\frac{q^2}{2\pi \I}\(\xi\log\frac{\I\xi}{\pi R\Lambda\zeta}-\xi\)
- \frac{q}{2\pi} \sum_{m\ne 0}  \frac{1}{m} \,e^{-\I m q \xi}
\Ei\left( \I m q \xi \right) \, .
\ee

\subsubsection*{Regular complex Darboux coordinates and transition functions}

The above analysis motivates the following construction of
regular complex Darboux coordinates on the twistor space $\cZ_\cS$.
First, we introduce two functions
\be
\cH^{\pm}(\xi)  \equiv
\frac{1}{2\pi\I} \sum_{m=1}^{\infty}  \frac{1}{m^2}
e^{\pm\I m q \xi}
\Ei\left( \mp\I m q \xi \right)
=\frac{1}{2\pi\I}\int_0^{\pm\I\infty} \frac{\de\xi'}{\xi-\xi'}\, \Li_2\(e^{\pm \I q \xi'}\)\ .
\label{tranfunreg}
\ee
Moreover, we consider a four-patch covering of  $\CP$ (see Fig.\ \ref{fig_sphere}, left):
the first patch $\cU_+$ surrounds the north pole
and extends along the contours $\ell_\pm$ down to the equator. The second patch $\cU_-$ surrounds
the south pole and similarly  extends halfway along $\ell_\pm$,  with a non-vanishing
intersection with $\cU_+$. The rest of  $\CP$ consists of two connected
parts belonging to $V_>$ and $V_<$ defined above, covered by two patches
$\cU_0$ and $\cU_{0'}$
which overlap with $\cU_+$ and $\cU_-$ but stay away from the contours $\ell_\pm$.

To this covering we associate the following transition functions
\be
\begin{split}
\Hij{0+} &=\Hij{0'+}=-\frac{q^2}{4\pi \I}\(\xi^2\log\frac{\I\zeta\xi}{\pi R\Lambda}-\frac{3}{2}\,\xi^2\)
+\cH^{+}(\xi)+\cH^{-}(\xi)\, ,
\\
\Hij{0-} &=\Hij{0'-}=\frac{q^2}{4\pi \I}\(\xi^2\log\frac{\I\xi}{\pi R\Lambda\zeta}-\frac{3}{2}\,\xi^2\)
-\cH^{+}(\xi)-\cH^{-}(\xi)\, .
\end{split}
\label{transGMN}
\ee
The momentum coordinates can then be obtained using the general
result eq.\ (3.38) of \cite{Alexandrov:2008ds} (adapting notations),
\be
\txii{i} (\zeta)=
\vrh +\sum_{j} \oint_{C_j} \frac{\de\zeta'}{2\pi \I\,\zeta'}\,
\frac{\zeta+\zeta'}{2 (\zeta'-\zeta)}\,
 \p_{\xi}H^{[0j]}(\zeta')\, ,
\label{mui}
\ee
which ensures that $\txii{0}$ (resp. $\txii{0'}$) is regular in $\cU_0$
(resp. $\cU_{0'}$), while $\txii{\pm}$ are regular in $\cU_\pm$.
Picking up the residues at $\zeta = 0$ and $\zeta = \infty$ in $\cU_+$ and $\cU_-$,
respectively, it is straightforward to check that the first term in
\eqref{transGMN} reproduces the weak coupling result \eqref{txiGMN} with $\delta\txi=0$,
upon identifying $\vrh=\theta_m -\frac{q^2}{4\pi \I}\,\theta_e\log (a/\ba)$.

\EPSFIGURE{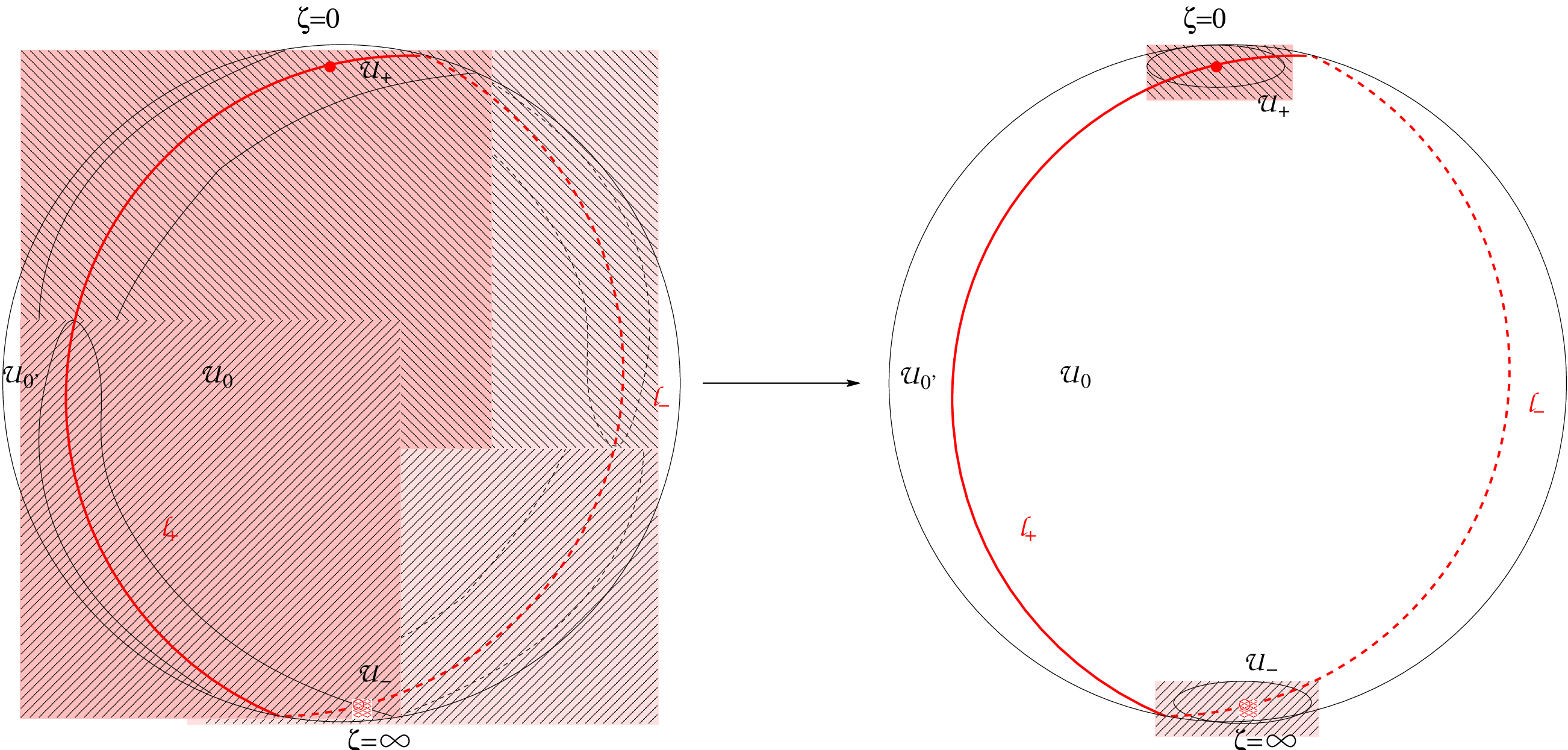,height=7.5cm}{Two coverings of $\CP$. The
covering on the left, described above \eqref{transGMN},
allows to reduce \eqref{mui} to figure-eight
contours around 0 and $\infty$. The covering on the right
is obtained in the limit where the strips $\cU_\pm$ go to
zero width along the meridians $\ell_\pm$, while maintaining a non-zero size at
the north and south pole. \label{fig_sphere}}

To see that the last two terms in \eqref{transGMN} reproduce $\delta\txi$,
note that
\be
\p_\xi\cH^\pm=\mp\frac{q}{2\pi}
\int_0^{\pm\I\infty} \frac{\de\xi'}{\xi-\xi'} \,\log\(1-e^{\pm \I q \xi'}\)
+\frac{\pi}{12\I\xi}\, .
\label{dertransGMN}
\ee
This representation makes it apparent that $\cH^\pm$ has
two logarithmic cuts  in the $\zeta$-plane, extending
from the north and the south poles inside $\cU_\pm$, respectively,
to the two zeros of $\xi$ inside $\cU_0\cup\cU_{0'}$.
Due to
$
\Hij{0+}_{\rm inst} = - \Hij{0-}_{\rm inst},
$
this situation is analogous to the one described in Section 3.4 of \cite{Alexandrov:2008ds}.
By a similar analysis one may conclude that the corresponding contribution to
$\txii{0}$ (and $\txii{0'}$)
is given by the sum of two integrals of $\p_\xi\cH^\pm$ along ``figure-eight"
contours encircling $\zeta=0$ and $\zeta=\infty$.
Since the first term in \eqref{dertransGMN} vanishes at these points,
there are no contributions from the poles in the measure, and
these integrals can be reduced to the integrals along $\ell_\pm$
of the discontinuity $\pm\I q \log\(1-e^{\pm \I q \xi}\)$ of $\p_\xi\cH^\pm$ across the cut,
reproducing $\delta\txi$ in \eqref{deltaxi}. Moreover, the last term in \eqref{dertransGMN}
does not contribute since $(\zeta\xi)^{-1}$ is regular in $\cU_\pm$.
Thus, the coordinates $\xii{0},\txii{0}$  agree with the ones defined
in \cite{Gaiotto:2008cd} in the patch $\cU_0$, and the same is true in the patch $\cU_{0'}$.

It is perhaps useful to note that, as depicted on Figure \ref{fig_sphere},
the patches $\cU_+$ and $\cU_-$ may be shrunk
to infinitesimal width along the contours $\ell_\pm$, while retaining a finite size
around the north and south pole, respectively. However, the fact that
the transition function $\Hij{00'}=\Hij{0+}+\Hij{+0'}$
vanishes does not imply that the coordinates $\txi$ are continuous along
$\ell_\pm$: indeed, the transition function $\Hij{0+}$ and $\Hij{+0'}$ have a discontinuity
along $\ell_\pm$, which reproduces the shifts  \eqref{disc1} and \eqref{disc2} in the
process of analytic continuation.

\subsection{Extension to QK and contact geometry  \label{ap_rig}}

It is now straightforward to apply the above construction to D-instanton corrections
to the hypermultiplet branch, since the twistor line $\txi(\zeta)$ \eqref{txiGMN} is
essentially given by the same integral as the one appearing in \eqref{IKkl}.
In this way, we can formulate the results of  Section \ref{sec:Dinst} in a rigorous way, avoiding
the use of open contours.

To this end, we use the same four patch covering as in the previous
subsection (cf.\ Fig. \ref{fig_sphere}), with the transition functions
\be
\label{gensymp2}
\begin{split}
\hHij{0+}&=\hHij{0'+}=  -\frac{\I}{2}\, F(\xi)
+\frac{1}{(2\pi)^3}
\ {\sum\limits_{\scriptsize\lefteqn{(k_\Lambda,l^\Lambda)_+}}}'\ \hnkl
\int_0^{-\I\infty} \frac{\Xi\,\de\Xi}{\Xikl^2-\Xi^2}\, \Li_2\left(e^{-2\pi \I\, \Xi} \right)\, ,
\\
\hHij{0-}& =\hHij{0'-}=   -\frac{\I}{2}\, \bF(\xi)
-\frac{1}{(2\pi)^3}
\ {\sum\limits_{\scriptsize\lefteqn{(k_\Lambda,l^\Lambda)_+}}}'\ \hnkl
\int_0^{-\I\infty} \frac{\Xi\,\de\Xi}{\Xikl^2-\Xi^2}\, \Li_2\left(e^{-2\pi \I\, \Xi} \right)\, ,
\end{split}
\ee
where
\be
\Xikl=k_\Lambda \xi^\Lambda-l^\Lambda \rho_\Lambda\, ,
\ee
and the only non-vanishing anomalous dimension is $\ci{+}_\alpha= \chi_X/(96\pi)$.
This can be shown by the same line of reasoning as below \eqref{dertransGMN}.
The description based on the function \eqref{prepH} is obtained in the limit where
$\cU_+$ and $\cU_-$ are shrunk to infinitesimal width along the contours $\ell_\pm$.

It would be interesting to understand the analytic
structure of the twistor lines at $\varpi=0,\infty$,
and thereby to extend the four patch construction described here,
beyond the leading instanton approximation.

%\bibliography{combined}
%\bibliographystyle{utphys}

\providecommand{\href}[2]{#2}\begingroup\raggedright\endgroup

\end{document}